\documentclass[%
reprint,
superscriptaddress,
amsmath,amssymb,
aps,
pra,
floatfix,
longbibliography
]{revtex4-2}

\usepackage{graphicx} 
\usepackage{dcolumn} 
\usepackage{bm} 
\usepackage{bbm}
\usepackage{braket}
\usepackage{comment}
\usepackage[dvipsnames]{xcolor}

\usepackage{booktabs}

\newcommand\identity{1\kern-0.25em\text{l}}

\definecolor{darkblue}{rgb}{0.,0.,0.6}
\definecolor{darkgreen}{rgb}{0.,0.6,0.0}
\definecolor{darkorange}{rgb}{0.8,0.4,0.2}

\usepackage[colorlinks=true, citecolor=blue, linkcolor=blue, urlcolor=blue]{hyperref}

\begin{document}
\preprint{APS/123-QED}
\raggedbottom

\title{Correlated Pair Creation with Positive- and Negative-Frequency Oscillators in a Trapped-Ion Quantum Simulator}

\author{Visal So}
\thanks{Present Address: California Institute of Technology, Pasadena, CA 91125}
\email{visalso@caltech.edu}
\affiliation{Department of Physics and Astronomy and Smalley-Curl Institute, Rice University, Houston, TX 77005, USA}
\author{Mingjian Zhu}
\affiliation{Department of Physics and Astronomy and Smalley-Curl Institute, Rice University, Houston, TX 77005, USA}
\author{Han Pu}
\affiliation{Department of Physics and Astronomy and Smalley-Curl Institute, Rice University, Houston, TX 77005, USA}
\author{Guido Pagano}
\affiliation{Department of Physics and Astronomy and Smalley-Curl Institute, Rice University, Houston, TX 77005, USA}

\date{\today}

\begin{abstract}

Spin-dependent forces enable independent control over the effective frequencies of bosonic motional modes and their coupling strengths to spins in trapped-ion spin-boson quantum simulators. We show that the relative signs of these effective frequencies provide an important programmable resource for engineering distinct multimode dynamics. Focusing on a two-mode quantum Rabi model, changing the frequencies of the two modes from the same sign to opposite signs reorganizes the resonance structure of the spin-mediated bosonic interaction, converting the dominant resonant process from excitation exchange to correlated pair creation. In the dispersive regime, the resulting effective interaction realizes a two-mode parametric amplifier with stable oscillatory, critical, and exponentially amplifying regimes separated by a spin-dependent instability threshold. Across these regimes, the dynamics generate collective-quadrature squeezing and bipartite motional entanglement. Preparing the spin in a coherent superposition further allows different spin components to experience distinct stability regimes, generating tripartite spin-boson entanglement, while spin postselection can coherently combine the conditional bosonic branches to generate non-Gaussian entangled bosonic states. Finally, we outline an implementation using parameters within reach of current trapped-ion platforms.
\end{abstract}

\maketitle

\section{Introduction}

Spin-boson models provide a fundamental setting for investigating the interplay between discrete quantum degrees of freedom and bosonic excitations and arise throughout quantum optics, condensed-matter physics, and chemical dynamics. A paradigmatic example is the quantum Rabi or single-mode spin-boson model, which describes a two-level system coupled to a harmonic oscillator. Despite its minimal structure, the model is integrable~\cite{braak2011integrability} and exhibits rich spin-boson dynamics, including a quantum phase transition associated with the loss of stability of the normal configuration in the classical oscillator limit~\cite{hwang2015qpt}. Generalizations to multiple bosonic modes introduce additional collective phenomena because the spin can mediate interactions between otherwise independent oscillators, allowing the character and stability of the resulting bosonic dynamics to be controlled through the underlying spin-boson coupling~\cite{mariantoni2008switch,shen2021twomoderabi,mcaleese2025engineering,chilingaryan2015twomode}.

A central distinction in coupled systems with two bosonic modes is between excitation exchange and pair creation. With $a$ and $b$ denoting the annihilation operators of the two bosonic modes, respectively, exchange interactions of the form $a^\dagger b+ab^\dagger$ coherently transfer excitations between modes and underlie beam-splitter operations and mode hybridization~\cite{campos1989beamsplitter,zhang2019bilinear,gan2020hybrid,katz2023programmable}. In contrast, interactions of the form $ab+a^\dagger b^\dagger$ create and annihilate excitations in correlated pairs and form the basis of non-degenerate parametric amplification and two-mode squeezing~\cite{caves1985twophoton1,schumaker1985twophoton2}. Starting from vacuum, such interactions generate two-mode-squeezed states characterized by enhanced correlations between the modes and reduced fluctuations of appropriate collective quadratures. These states constitute an important resource for continuous-variable quantum information, entanglement generation, and quantum-enhanced sensing~\cite{duan2000inseparability,simon2000peres,metzner2024twomode}.

Closely related pair-creation physics appears in interacting atomic many-body systems. In spinor Bose-Einstein condensates, spin-exchange collisions convert pairs of atoms from one internal state into correlated pairs occupying two other spin states, providing a matter-wave analogue of parametric down-conversion~\cite{law1998quantum}. Experiments have demonstrated amplification of vacuum fluctuations through such spin-mixing processes~\cite{klempt2010parametric}, strong quantum correlations between the generated atom pairs~\cite{bookjans2011strong}, continuous-variable entanglement of twin-atom modes~\cite{gross2011atomic}, and the generation and characterization of a two-mode squeezed vacuum~\cite{qu2020probing}. More recently, two-mode squeezing has been directly engineered between distinct motional modes of a trapped ion using parametrically driven trap potentials~\cite{metzner2024twomode}. These realizations illustrate the broad relevance of correlated pair production as a mechanism for generating nonclassical bosonic states.

Pair interactions are also closely connected to dynamical instability. For coherently coupled modes with the same energy signature, a near-resonant exchange interaction typically produces level repulsion and coherent mode hybridization~\cite{bernier2018level}. The situation changes when modes of opposite energy signature are coupled. In this case, excitations can be created in pairs while approximately conserving the total energy because the positive contribution of one mode is compensated by the negative contribution of the other~\cite{jain2007quantum,kohler2018negative}. As the coupling becomes sufficiently strong relative to the residual frequency mismatch, the normal-mode frequencies can coalesce and become dynamically unstable, converting bounded oscillatory motion into exponentially amplified fluctuations~\cite{bernier2018level,kohler2018negative}. Such behavior provides a dynamical counterpart to equilibrium critical phenomena, in which the loss of stability of a reference configuration manifests as amplification of fluctuations in time rather than stabilization around a new equilibrium configuration.

Experimentally, trapped ions provide a controllable platform for studying these concepts through multimode quantum Rabi models, with long-lived internal states serving as effective spins and collective motional modes as bosons. Laser-induced spin-dependent forces coherently couple these degrees of freedom~\cite{leibfried2003ion,molmer1999thermal,molmer2000entanglement,haljan2005sdf,kim2009spin,monroe2021programmable,pagano2025varenna} and have already enabled theoretical proposals and experimental realizations of single-mode quantum Rabi physics from weak to deep-strong coupling~\cite{pedernales2015quantumrabi,lv2018rabi,cai2021rabi}, as well as multimode spin-boson simulations of conical intersections, charge transfer, and excitation transfer~\cite{valahu2023conical,whitlow2023conical,gorman2018VAET,so2024electrontransfer,sun2025quantum,so2025multimode,navickas2025chemical}. In these driven systems, the detuning from a motional sideband defines an effective oscillator frequency in a rotating frame~\cite{schneider2012spinboson,pedernales2015quantumrabi}, allowing the spin-boson coupling to become comparable to or exceed the simulated mode frequency~\cite{lv2018rabi,cai2021rabi}. Crucially, this frequency has a controllable sign. Reversing it changes the direction of rotating-frame phase-space evolution without destabilizing the physical motional mode. The presence of relative signs in multimode systems can then modify the resonance conditions of spin-mediated bosonic processes, providing an additional programmable control parameter for trapped-ion spin-boson simulation.

Although spin-dependent forces can routinely be generated on either side of a motional sideband, previous trapped-ion spin-boson simulations, including Refs.~\cite{lv2018rabi,cai2021rabi,so2024electrontransfer,so2025multimode}, have primarily employed positive effective frequencies. Here, we exploit the relative signs of effective bosonic frequencies to engineer mixed-sign spin-boson dynamics. By using a spin linearly coupled to two bosonic modes as a minimal extension of the quantum Rabi model, we show that controlling the frequency signs changes the resonant spin-mediated interaction from excitation exchange to pair creation. In the dispersive regime, coupling positive- and negative-energy modes produces a dynamical instability whose resonance condition and threshold depend on spin-conditioned dispersive shifts. Within a fixed spin sector, the resulting dynamics generate two-mode squeezing and can violate the Duan-Simon separability bound.

Beyond mediating bosonic interactions, a coherent spin superposition renders the bosonic dynamics branch dependent, realizing distinct stability regimes across different spin branches and generating tripartite spin-boson entanglement. Measurement in a non-$\sigma_z$ basis can then induce interference between these branches to produce non-Gaussian entangled bosonic states whose logarithmic negativity remains finite even when the second-moment Duan-Simon inequality is not violated. We also discuss the compatibility of the proposed control requirements with existing trapped-ion quantum simulators. More broadly, we demonstrate that mixed-sign frequencies provide a programmable control parameter for engineering nonequilibrium bosonic dynamics through spin-mediated interactions without an explicit bosonic drive.

The remainder of the paper is organized as follows. Secs.~\ref{sec:hamiltonianengineering} and \ref{sec:meaning} introduce the implementation and physical interpretation of signed effective mode frequencies. Sec.~\ref{sec:hybrid} develops the mixed-sign two-mode model, including spin-mediated pair dynamics, spin-conditioned entanglement, and spin-postselected non-Gaussian states. Finally, Sec.~\ref{sec:experiment} discusses an experimental implementation in trapped ions.

\section{Effective mode frequencies}
\label{sec:synthetic}

\subsection{Hamiltonian engineering}
\label{sec:hamiltonianengineering}

\begin{figure}
    \centering
    \includegraphics[width=1\linewidth]{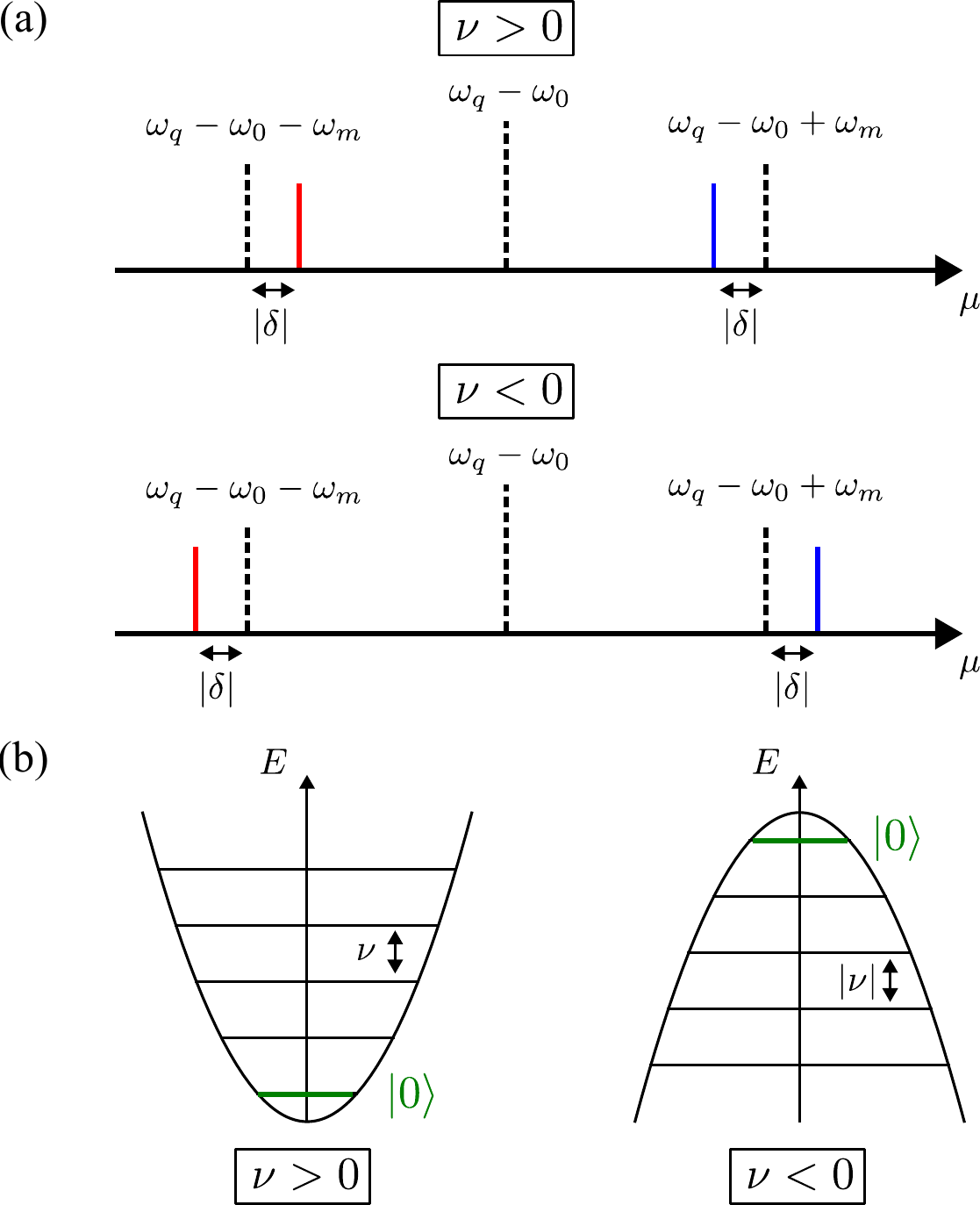}
    \caption{Engineering the sign of an effective mode frequency. (a) The two frequency components of the bichromatic laser field [colored vertical bars] are placed symmetrically about the shifted carrier $\omega_q-\omega_0$ at a distance $\omega_m+\delta$. For $\delta<0$, the tones lie between the carrier and the motional sidebands and realize an ordinary oscillator with $\nu=-\delta>0$. For $\delta>0$, they lie outside the sidebands and realize a negative-frequency oscillator with $\nu=-\delta<0$. (b) Corresponding energy ladders in the rotating frame. For $\nu<0$, the ladder is inverted, and the phonon vacuum $\ket{0}$ lies at the top of the effective energy spectrum. We emphasize that the parabolic curves depict rotating-frame energies rather than laboratory trapping potentials, and the underlying physical motional modes remain stable positive-frequency oscillators.}
    \label{fig:scheme}
\end{figure}

Throughout this work, we set $\hbar=1$. We consider a spin-$1/2$ degree of freedom, with basis states $\{\ket{\uparrow},\ket{\downarrow}\}$, encoded in two long-lived internal states of one ion in a Coulomb crystal with energy splitting $\omega_q$. This effective spin is coupled to a collective motional mode of frequency $\omega_m$ with the Lamb-Dicke parameter $\eta$ using a bichromatic laser field. As shown in Fig.~\ref{fig:scheme}(a), the bichromatic frequencies are chosen as
\begin{equation}
  \mu_{\pm} = (\omega_q-\omega_0)\pm(\omega_m+\delta),
  \label{eq:tones}
\end{equation}
where $\mu_+$ and $\mu_-$ address the blue and red sidebands, respectively, and the two tones are driven with equal Rabi frequency $\Omega$. Here, $-\omega_0$ is a deliberately introduced offset from the qubit resonance.

We work in a rotating frame generated by
\begin{equation}
  H_{\rm ref}=\frac{\omega_q-\omega_0}{2}\sigma_z+(\omega_m+\delta)a^\dagger a .
  \label{eq:Href}
\end{equation}
In this frame, the bare spin and motional Hamiltonian becomes
\begin{equation}
  \frac{\omega_q}{2}\sigma_z+\omega_m a^\dagger a-H_{\rm ref}=\frac{\omega_0}{2}\sigma_z-\delta\,a^\dagger a .
\end{equation}
Within the Lamb-Dicke regime, $\eta\sqrt{\braket{(a+a^\dagger)^2}}\ll1$, and after the optical rotating-wave approximation in the sideband-resolved regime ($|\omega_0|,|\delta|,\Omega\ll\omega_m$), the near-resonant red- and blue-sideband couplings become time-independent~\cite{schneider2012spinboson,pedernales2015quantumrabi,lv2018rabi,so2024electrontransfer,pagano2025varenna}. The resulting Hamiltonian is
\begin{equation}
H=\frac{\omega_0}{2}\sigma_z+\nu a^\dagger a
+g\sigma_\phi\left(ae^{-i\psi}+a^\dagger e^{i\psi}\right),
\label{eq:genqrm}
\end{equation}
where $\nu\equiv-\delta$, $g=\eta\Omega/2$, and $\sigma_\phi\equiv\sigma_x\cos\phi-\sigma_y\sin\phi$. The effective spin and motional phases, $\phi$ and $\psi$, are determined by the optical phases of the two frequency tones. For $\phi=\psi=0$, Eq.~\eqref{eq:genqrm} realizes the quantum Rabi model with independently programmable parameters,
\begin{equation}
H_{\rm QRM}(\omega_0,\nu,g)=\frac{\omega_0}{2}\sigma_z+\nu a^\dagger a
+g\sigma_x\left(a+a^\dagger\right).
\label{eq:qrm}
\end{equation}
Here, the effective qubit splitting $\omega_0$ is set by the common offset of the two tones from the qubit transition, the spin-boson coupling $g=\eta\Omega/2$ is controlled by the optical power, and the mode frequency $\nu=-\delta$ is set by the sideband detuning~\cite{pedernales2015quantumrabi,lv2018rabi}. With the convention of Eq.~\eqref{eq:tones}, $\delta<0$ produces a positive oscillator frequency with $\nu>0$, whereas $\delta>0$ produces a negative oscillator frequency with $\nu<0$ (see Fig.~\ref{fig:scheme}(b) and Ref.~\footnote{It is important to note that the bookkeeping sign of the effective frequency is tied to the convention used to parametrize the tone placement. If the bichromatic frequencies in Eq.~\eqref{eq:tones} are instead written as $\mu_\pm = (\omega_q - \omega_0) \pm (\omega_m - \delta)$, the same physical tone placement corresponds to $\delta \to -\delta$, and the rotating-frame oscillator term becomes $+\delta\, a^\dagger a$, reversing the relation between the sign of the detuning and the sign of $\nu$. However, the physical content is still embedded in the resulting dynamics rather than in the bookkeeping choice.}). Under the term $\nu a^\dagger a$, the Heisenberg-picture mode operator evolves as $a(t) = a(0)e^{-i\nu t}$, so opposite signs of $\nu$ correspond to opposite senses of phase-space rotation relative to the drive-defined phase reference, providing a direct calibration signature. Negative-frequency motional modes can also arise naturally in ion systems based on Penning traps, where, in the normal-mode description, the magnetron branch carries a negative mode frequency~\cite{jain2020scalable}. By contrast, in our approach, the effective frequency is synthetically programmed through the sideband detuning in the rotating frame of an otherwise stable positive-frequency motional mode, allowing direct control over both its magnitude and sign.

The generalization to several modes is rather straightforward. If the motional modes are sufficiently spectrally resolved that unintended off-resonant couplings can be neglected, then, within the same Lamb-Dicke and sideband-resolved approximations, one may simultaneously use multiple bichromatic tones. Each tone pair is detuned by $\delta_k$ from the corresponding red and blue sidebands of mode $k$ and produces a spin-phonon coupling strength $g_k$. This gives
\begin{equation}
H=\frac{\omega_0}{2}\sigma_z+\sum_k\left[\nu_k a_k^\dagger a_k+g_k\sigma_{\phi_k}\left(a_k e^{-i\psi_k}+a_k^\dagger e^{i\psi_k}\right)\right],
\label{eq:multimode}
\end{equation}
where $\nu_k=-\delta_k$. Thus, each effective mode frequency can, in principle, be programmed independently in both magnitude and sign, together with the corresponding spin-phonon coupling strength and optical phases. A mixed-sign pair of effective mode frequencies can alternatively be realized with a single bichromatic tone pair placed between two motional sidebands. In that case, the detunings from the two modes have opposite signs, but their amplitudes are not independently tunable because $\nu_2-\nu_1=\omega_{m,2}-\omega_{m,1}$. In addition, $g_k=\eta_k\Omega/2$ fixes the relative coupling strengths through the Lamb-Dicke parameters. Thus, the single-pair scheme provides an economical realization of two opposite-sign oscillators, whereas the multi-pair configuration of Eq.~\eqref{eq:multimode} is required for independent control of the effective mode frequencies and couplings.

Moreover, near-carrier drives on the qubit transition can be applied simultaneously with the sideband tones. In the rotating frame generated by Eq.~\eqref{eq:Href}, a drive with frequency $\mu_c=\omega_q-\omega_0$, Rabi frequency $\Omega_c$, and phase $\phi_c$ is time-independent and contributes $H_c=(\Omega_c/2)\sigma_{\phi_c}$. Such drives provide independent single-qubit control terms in addition to the effective splitting generated by the common sideband offset $\omega_0$. Alternatively, one may set the common sideband offset to zero and generate the effective qubit splitting using a resonant carrier drive, as in Refs.~\cite{so2024electrontransfer,so2025multimode,so2026finitetemp}. In that case, a carrier tone at $\mu_c=\omega_q$ contributes $(\Omega_c/2)\sigma_{\phi_c}$ in the qubit rotating frame. After a suitable spin-basis rotation, this term plays the role of the qubit-splitting term in the simulated Hamiltonian, with the effective splitting set by $\Omega_c$ rather than by $\omega_0$. Adding a second resonant carrier tone with a relative phase shift of $\pi/2$ generates an orthogonal single-qubit term. Together with spin-dependent forces written in the rotated basis, this realizes Hamiltonians of the form
\begin{equation}
  H_{\rm LVC}=\frac{\Delta E}{2}\sigma_z+V\sigma_x+\sum_k\left[\nu_k a_k^\dagger a_k+g_k\sigma_z(a_k+a_k^\dagger)\right],
  \label{eq:ETmodel}
\end{equation}
which was used in the multimode charge- and exciton-transfer experiments of Refs.~\cite{so2025multimode,so2026finitetemp}. Here, $\Delta E$ and $2V$ are the Rabi frequencies of the two carrier tones.

\subsection{Negative-frequency modes}
\label{sec:meaning}

When using the rotating-frame transformation generated by Eq.~\eqref{eq:Href} to realize negative-frequency models, several features define the physical content of the construction. First, the rotating-frame generator in Eq.~\eqref{eq:Href} commutes with the phonon-number operator $a^\dagger a$ and with $\sigma_z$. Therefore, phonon-number distributions, spin populations in the $\sigma_z$ basis, and correlation functions constructed from these observables have identical values in the laboratory and simulated frames. These quantities can thus be measured using standard techniques without additional frame reconstruction. By contrast, phase-sensitive motional observables are not invariant under the transformation, as the mode operator acquires a rotating-frame phase. Quadrature-sensitive measurements, such as displacement probes or characteristic-function measurements, must therefore use phases referenced to the same rotating frame, as is standard in trapped-ion motional-state measurements~\cite{fluhmann2019encoding,fluhmann2020direct}.

Second, the spectrum of $-|\nu|a^\dagger a$ is unbounded from below, but this does not indicate a laboratory-frame instability. The simulated Hamiltonian is the generator of dynamics in a driven rotating frame, not the fundamental energy of the trapped-ion crystal. In the laboratory frame, the motional mode remains an ordinary positive-energy oscillator with frequency $\omega_m$, so increasing the phonon number still increases the bare mechanical energy of the ion crystal. Energy conservation applies to the full system consisting of the ions and the applied optical fields, where changes in the rotating-frame energy are balanced by changes in the laboratory-frame ion energy together with work done by, or returned to, the drive. Thus, the negative frequency should be understood as a sign reversal of the harmonic term in the rotating-frame Hamiltonian rather than as a negative bare mechanical energy.

Finally, the negative-frequency oscillator can be rewritten as
\begin{equation}
H=-|\nu|a^\dagger a=-\frac{|\nu|}{2}\left(X^2+P^2\right)+\frac{|\nu|}{2},
\label{eq:negosci}
\end{equation}
where $a=(X+iP)/\sqrt{2}$. Up to the additive constant, this Hamiltonian maps to a stable classical oscillator with negative mass moving in an inverted harmonic potential. Therefore, it should be distinguished from the inverted parabolic-barrier oscillator, which is mathematically described by
\begin{equation}
H_{\rm inv}=\frac{P^2}{2}-\frac{|\kappa|X^2}{2}.
\label{eq:invertosci}
\end{equation}
The oscillator in Eq.~\eqref{eq:invertosci} has a continuous scattering spectrum and hyperbolic phase-space dynamics, with exponentially growing and decaying directions~\cite{sundaram2024inverted}. By contrast, the negative-frequency mode in Eq.~\eqref{eq:negosci} has a discrete, equally spaced ladder whose ordering is inverted in the rotating-frame energy (see Fig.~\ref{fig:scheme}(b)). Its free evolution remains stable with the evolution operator $e^{+i|\nu|t a^\dagger a}$. Physically, the mode undergoes phase-space rotation with the opposite sense from an ordinary oscillator. In isolation, it therefore exhibits no dynamical instability. As will be shown in the next section, exponential growth of the bosonic occupation can nevertheless arise when negative- and positive-frequency modes are coupled in certain parameter regimes.

\section{Positive- and negative-frequency mode pairs}
\label{sec:hybrid}

Spin-boson models with isolated negative-frequency oscillators are often directly related to their positive-frequency counterparts by spin rotations and, for suitable Hamiltonian structures, by an overall reversal of the closed-system Hamiltonian. Thus, reversing the sign of an isolated effective frequency does not necessarily produce physically distinct dynamics by itself (see Appendix~\ref{app:neg_rabi} for the discussion of the negative-frequency single-mode quantum Rabi model). Instead, it can be exploited operationally for time-reversal protocols. A discussion of its use for Loschmidt-echo benchmarking of spin-boson simulators is provided in Appendix~\ref{sec:diagnostic}. In contrast, when positive- and negative-frequency modes are present simultaneously, their relative signs modify the resonance conditions for interactions between the modes, producing dynamics that cannot generally be reduced to an overall reversal of the Hamiltonian evolution.

\subsection{Spin-mediated bosonic pair amplification}
\label{sec:paircreation}

To illustrate how these modified resonance conditions can convert spin-mediated excitation exchange into pair creation, we consider a two-mode quantum Rabi model in which a spin is coupled to two bosonic modes with opposite-sign frequencies, $\nu_a>0$ and $\nu_b<0$. The Hamiltonian is
\begin{align}
H_{\rm 2QRM}={}&\frac{\omega_0}{2}\sigma_z+\nu_a a^\dagger a+\nu_b b^\dagger b \nonumber\\
&+\sigma_x\left[g_a\left(a+a^\dagger\right)+g_b\left(b+b^\dagger\right)\right].
\label{eq:2qrm}
\end{align}
To derive the spin-mediated bosonic interaction, we apply a Schrieffer-Wolff transformation that perturbatively eliminates spin-changing processes~\cite{schrieffer1966anderson,bravyi2011schrieffer,hwang2015qpt}. For this purpose, we write $H_{\rm 2QRM}=H_0+V$, with
\begin{equation}
H_0=\frac{\omega_0}{2}\sigma_z+\nu_a a^\dagger a + \nu_b b^\dagger b, \quad V=\sigma_x \widetilde{X},
\end{equation}
where $\widetilde{X}=g_a\left(a+a^\dagger\right)+g_b\left(b+b^\dagger\right)$. We work in the dispersive regime, where the spin-flip transitions are far detuned from the bosonic sidebands,
\begin{equation}
\left|\omega_0\pm\nu_k\right|\gg|g_k|\sqrt{n_k+1},\quad k=a,b,
\label{eq:dispersiveconditions}
\end{equation}
for all appreciably populated bosonic Fock states over the time interval of interest. This condition ensures that spin-changing processes remain virtual and that the following Schrieffer-Wolff expansion is controlled. We further assume $\omega_0\gg|\nu_k|$ when expanding the Schrieffer-Wolff transformation in powers of $\nu_k/\omega_0$~\cite{schrieffer1966anderson,bravyi2011schrieffer,hwang2015qpt}.

We choose the first-order anti-Hermitian generator $S$ to satisfy $\left[S,H_0\right]=-V$, giving
\begin{align}
S=\sum_{k=a,b}g_k\bigg[&\sigma_+\left(\frac{k}{\omega_0-\nu_k}+\frac{k^\dagger}{\omega_0+\nu_k}\right)\nonumber\\
-&\sigma_-\left(\frac{k}{\omega_0+\nu_k}+\frac{k^\dagger}{\omega_0-\nu_k}\right)\bigg].
\label{eq:sw_generator}
\end{align}
Expanding in $\nu_k/\omega_0$ gives
\begin{eqnarray}
S&=&\frac{i\sigma_y}{\omega_0}\widetilde{X}+\frac{\sigma_x}{\omega_0^2}\sum_{k=a,b}\nu_k g_k(k-k^\dagger)\nonumber\\
&\quad&+\,\mathcal{O}\!\left(\frac{\nu_{\max}^2}{\omega_0^3}\sum_{k=a,b}|g_k|\left\|(k+k^\dagger)\right\|\right).
\label{eq:expanded_SW_generator}
\end{eqnarray}
where $\nu_{\rm max}=\max(\nu_a,|\nu_b|)$, and $\left\|\cdot\right\|$ denotes the operator norm restricted to the appreciably populated occupation sector. The Baker-Campbell-Hausdorff expansion of the transformed Hamiltonian to second order in the spin-boson couplings yields
\begin{align}
H_{\rm eff}&=e^S H_{\rm 2QRM}e^{-S}\nonumber\\
&=H_0+\frac{1}{2}[S,V]+\mathcal{O}\left(\frac{\left\|\widetilde{X}\right\|^3}{\omega_0^2}\right).
\end{align}
The second term in Eq.~\eqref{eq:expanded_SW_generator} gives the leading correction associated with the bosonic dynamics. Substituting the retained terms of Eq.~\eqref{eq:expanded_SW_generator} into the commutator with $V$ gives
\begin{equation}
\frac{1}{2}[S,V]
=\frac{\sigma_z}{\omega_0}\widetilde{X}^2+\left(\sum_{k=a,b}\frac{\nu_k g_k^2}{\omega_0^2}\right)\identity+\mathcal{O}\!\left(\frac{\nu_{\max}^2}{\omega_0^3}\left\|\widetilde{X}\right\|^2\right).
\label{eq:SW_commutator_expansion}
\end{equation}
Since the second term in Eq.~\eqref{eq:SW_commutator_expansion} is proportional to the identity, it only shifts the overall energy and can be omitted. Keeping the leading dynamical contribution therefore yields
\begin{eqnarray}
H_{\rm eff}&\approx&\frac{\omega_0}{2}\sigma_z+\nu_a a^\dagger a+\nu_b b^\dagger b\nonumber\\
&\quad&\,+\frac{\sigma_z}{\omega_0}\left[g_a(a+a^\dagger)+g_b(b+b^\dagger)\right]^2.
\label{eq:2qrmSW}
\end{eqnarray}
Eq.~\eqref{eq:2qrmSW} is accurate to second order in the dispersive expansion and to leading dynamical order in $\nu_{\max}/\omega_0$. A complementary position-dependent rotation, presented in Appendix~\ref{app:position_rotation}, shows that the quadratic interaction in Eq.~\eqref{eq:2qrmSW} is also the leading expansion of the nonlinear dressed-spin potential.

Moving to the interaction picture generated by the free bosonic Hamiltonian,
$H_{\rm b}=\nu_a a^\dagger a+\nu_b b^\dagger b$, gives $a(t)=a e^{-i\nu_a t}$ and $b(t)=b e^{-i\nu_b t}$. Since $\nu_a>0$ and $\nu_b<0$, the two modes rotate in opposite directions in phase space. Projecting onto a fixed spin sector satisfying $\sigma_z\ket{s}=s\ket{s}$, with $s=\pm1$, the single-mode contributions produce dispersive shifts of the signed mode frequencies,
\begin{equation}
    \widetilde{\nu}_k=\nu_k+ \frac{2s g_k^2}{\omega_0}.
\end{equation}
They also generate single-mode squeezing terms $k^2+k^{\dagger 2}$ that rotate at frequencies $2\nu_k$ with amplitudes of order $g_k^2/\omega_0$.

The key distinction from the equal-sign case ($\nu_a, \nu_b>0$) arises from the cross term $2g_ag_b(a+a^\dagger)(b+b^\dagger)$ in Eq.~\eqref{eq:2qrmSW}. In the interaction picture, the pair operators evolve as
\begin{equation}
ab\rightarrow ab\,e^{-i(\nu_a+\nu_b)t},\quad
a^\dagger b^\dagger\rightarrow a^\dagger b^\dagger e^{+i(\nu_a+\nu_b)t},
\end{equation}
whereas the exchange operators evolve as
\begin{equation}
a^\dagger b\rightarrow a^\dagger b\,e^{+i(\nu_a-\nu_b)t},
\quad
ab^\dagger\rightarrow ab^\dagger e^{-i(\nu_a-\nu_b)t}.
\end{equation}
Thus, for opposite-sign frequencies, the pair terms can be near resonance when $\nu_a\approx|\nu_b|$, whereas the exchange terms generally rotate rapidly at
$\nu_a-\nu_b=\nu_a+|\nu_b|$.

After absorbing the secular dispersive shifts into the dressed mode frequencies and retaining the near-resonant pair terms, the effective bosonic Hamiltonian in spin sector $s$ for the effective Hamiltonian in Eq.~\eqref{eq:2qrmSW} becomes
\begin{eqnarray}
    H_{\rm RWA}^{(s)}&=&\widetilde{\nu}_a a^\dagger a + \widetilde{\nu}_b b^\dagger b + s\lambda\left(ab+a^\dagger b^\dagger\right)\nonumber\\
    &=& \frac{\widetilde{\nu}_a+\widetilde{\nu}_b}{2}(n_a+n_b)+\frac{\widetilde{\nu}_a-\widetilde{\nu}_b}{2}(n_a-n_b)\nonumber\\
    &\quad&\quad + s\lambda\left(ab+a^\dagger b^\dagger\right),
    \label{eq:detuned_pair_hamiltonian}
\end{eqnarray}
where $n_a=a^\dagger a$ and $n_b=b^\dagger b$ denote the number operators of modes $a$ and $b$, respectively, and $\lambda=(2g_ag_b)/\omega_0$. A sufficient set of conditions for this rotating-wave approximation is
\begin{equation}
2|\widetilde{\nu}_k|\gg\frac{g_k^2}{\omega_0},
\quad |\widetilde{\nu}_a-\widetilde{\nu}_b|\gg|\lambda|,
\end{equation}
so that the single-mode-squeezing and exchange terms remain
off resonant, while the dressed pair detuning,
\begin{equation}
    \Delta\widetilde{\nu}\equiv\widetilde{\nu}_a+\widetilde{\nu}_b=\nu_a+\nu_b+\frac{2s(g_a^2+g_b^2)}{\omega_0},
\end{equation}
may be comparable to $|\lambda|$.

\begin{figure}
    \centering
    \includegraphics[width=1\linewidth]{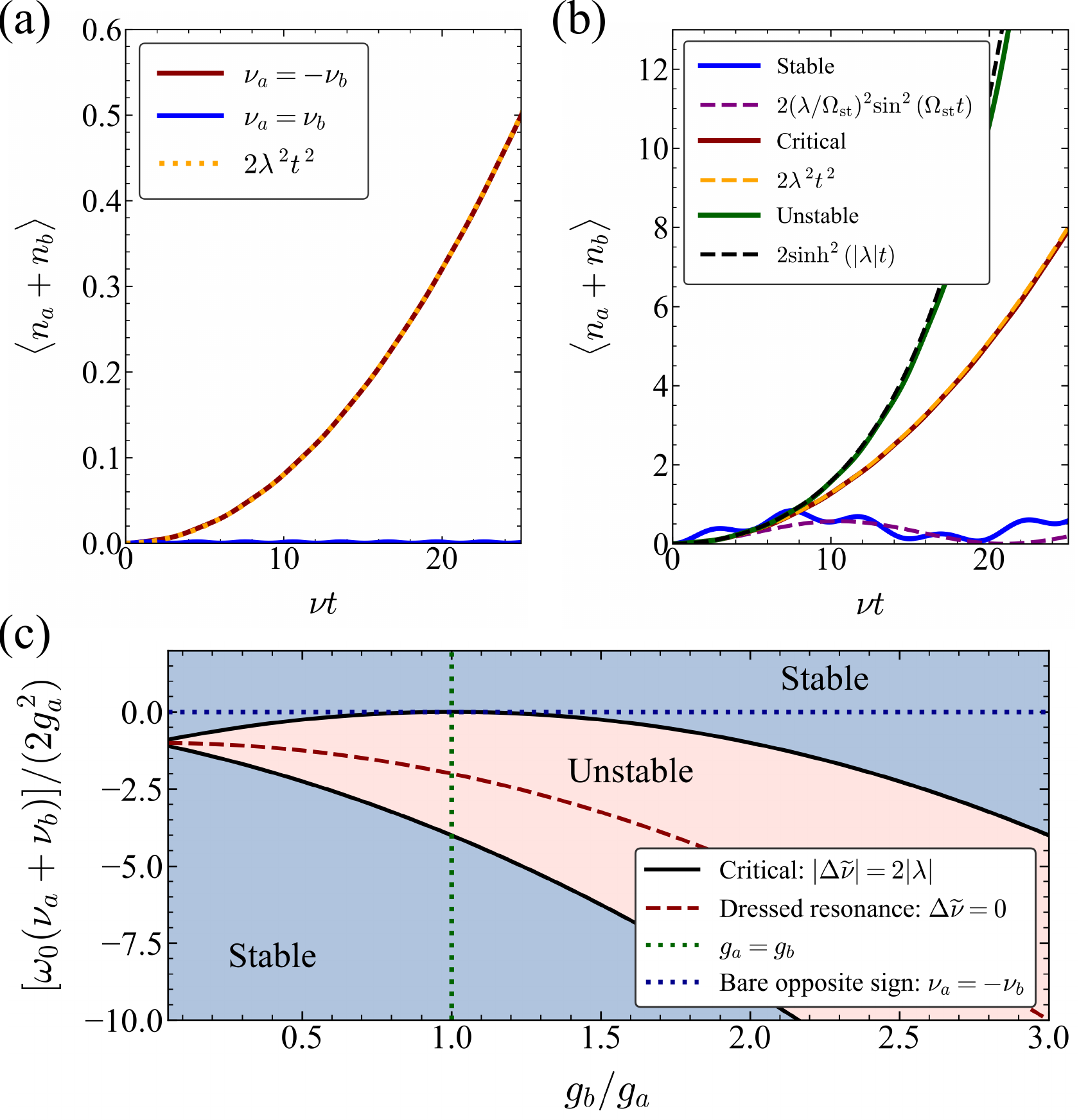}
    \caption{Bosonic pair dynamics and dynamical instability in the two-mode quantum Rabi model in the Schrieffer-Wolff frame. (a) Comparison between the all-positive-frequency case, $\nu_b=\nu$ (blue), and the opposite-sign case, $\nu_b=-\nu$ (red), for $g_a=g_b=\nu$. The opposite-sign case has $\lambda=0.02\nu$ and $\Delta\widetilde{\nu}=0.04\nu=2|\lambda|$, placing it at the critical boundary. (b) Stable, critical, and dynamically unstable regimes at fixed pair coupling $\lambda=0.08\nu$. The stable case (blue) has $g_a=\nu$, $g_b=4\nu$, $\nu_b=-\nu$, and $\Delta\widetilde{\nu}=0.34\nu$. The critical case (red) has $g_a=g_b=2\nu$, $\nu_b=-\nu$, and $\Delta\widetilde{\nu}=0.16\nu=2|\lambda|$. The unstable case (green) has $g_a=g_b=2\nu$, $\nu_b=-1.16\nu$, and $\Delta\widetilde{\nu}=0$. In panels (a) and (b), the system is initialized in the two-mode vacuum with $s=+1$, $\nu_a=\nu$, and $\omega_0=100\nu$. (c) Dynamical stability diagram predicted by the effective Hamiltonian as a function of the normalized bare frequency mismatch and coupling ratio. The diagram is shown for $s=+1$; the $s=-1$ diagram is obtained by reflection about the bare opposite-sign line $\nu_a=-\nu_b$. The stability diagram applies while the dispersive condition in Eq.~\eqref{eq:dispersiveconditions} remains satisfied.}
    \label{fig:paircreation}
\end{figure}

When the dressed pair-resonance condition $\Delta\widetilde{\nu}\approx0$ is satisfied, the effective Hamiltonian can be written as
\begin{equation}
H_{\rm RWA}^{(s)}\approx\frac{\widetilde{\nu}_a-\widetilde{\nu}_b}{2}
\left(n_a-n_b\right)+s\lambda\left(ab+a^\dagger b^\dagger\right).
\label{eq:effective_hamiltonian_after_rwa}
\end{equation}
The number-difference term  $n_a-n_b$ commutes with the pair interaction and vanishes within the $n_a=n_b$ subspace generated from the two-mode vacuum. The occupation dynamics is therefore governed by
\begin{equation}
H_{\rm pair}=s\lambda\left(ab+a^\dagger b^\dagger\right).
\label{eq:tms}
\end{equation}
This quadratic interaction is equivalent, up to phase conventions, to the standard nondegenerate-parametric-amplifier or two-mode-squeezing Hamiltonian~\cite{caves1985twophoton1,schumaker1985twophoton2}. Related interactions between distinct trapped-ion motional modes have been directly realized by modulating the trapping potential at the sum of the laboratory-frame mode frequencies~\cite{metzner2024twomode}.

For two modes with equal positive frequencies, the situation is reversed. The beam-splitter interaction $a^\dagger b+a b^\dagger$ is near resonance when $\widetilde{\nu}_a - \widetilde{\nu}_b\approx 0$, whereas the pair terms rotate at $\nu_a+\nu_b$ and average away~\cite{mariantoni2008switch,reuther2010dissipative,baust2015tunable}. Flipping the sign of one frequency, therefore, converts resonant mode exchange into two-mode squeezing. When $\Delta\widetilde{\nu}\approx0$, pair creation conserves the dressed signed free-mode energy even as the total bosonic occupation grows, realizing the parametric instability associated with level attraction, negative-mass instabilities, and positive- and negative-energy-mode mixing~\cite{jain2007quantum,bernier2018level,kohler2018negative}.

For an initial two-mode vacuum, the dynamics of Eq.~\eqref{eq:detuned_pair_hamiltonian} are governed by the dressed pair detuning $\Delta\widetilde{\nu}$ and the pair coupling strength $\lambda$. As derived in Appendix~\ref{app:pair_amplification}, the system is dynamically unstable when $|\lambda|>\frac{1}{2}\left|\Delta\widetilde{\nu}\right|$ with growth rate $\Gamma_{\rm inst}=\sqrt{\lambda^2-\Delta\widetilde{\nu}^2/4}$. On exact dressed resonance, $\Delta\widetilde{\nu}=0$, the mode occupations grow as
\begin{equation}
    \braket{n_a(t)}=\braket{n_b(t)}=\sinh^2(|\lambda| t).
    \label{eq:resonance_phonon_growth}
\end{equation}
Below the instability threshold, the pair dynamics are bounded and oscillatory, with frequency $\Omega_{\rm st}=\sqrt{\Delta\widetilde{\nu}^2/4-\lambda^2}$. At the critical boundary, $|\lambda|=\frac{1}{2}\left|\Delta\widetilde{\nu}\right|$, the exponential growth rate vanishes, and the vacuum populations instead grow algebraically as
\begin{equation}
\braket{n_a(t)}=\braket{n_b(t)}=\lambda^2t^2.
\end{equation}

\begin{figure*}
    \centering
    \includegraphics[width=1\linewidth]{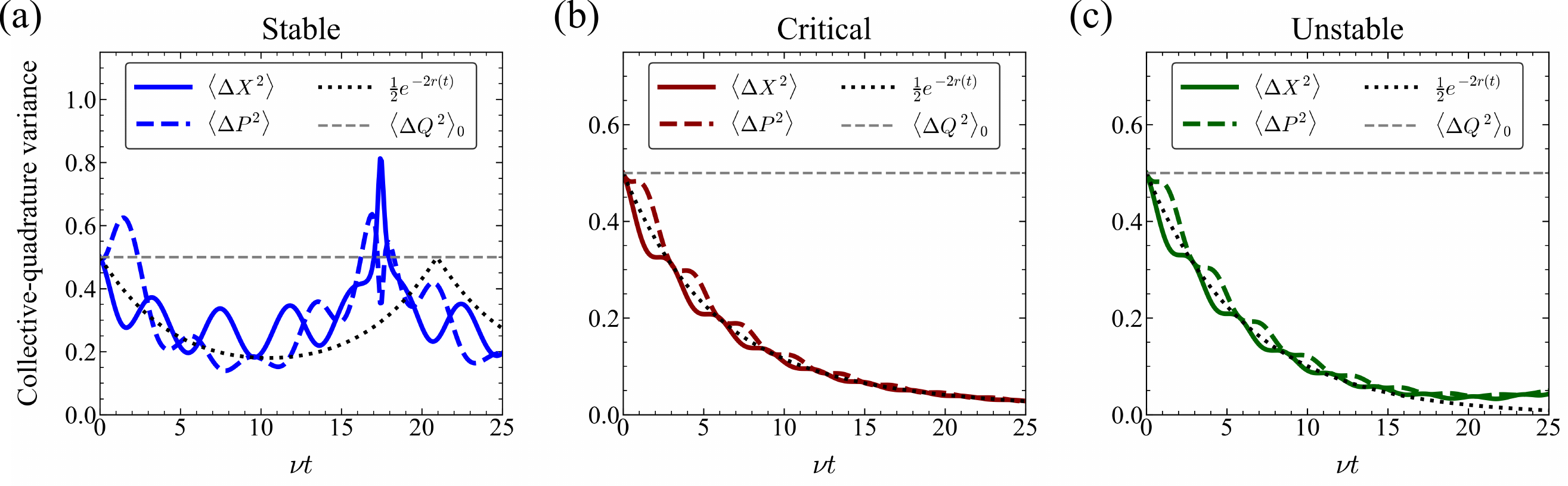}
    \caption{Collective-quadrature squeezing in the stable, critical, and dynamically unstable regimes. The initial state and parameter sets are the same as in Fig.~\ref{fig:paircreation}(b). Solid and dashed colored curves show the variances $\braket{\Delta X^2}$ and $\braket{\Delta P^2}$, respectively, for the full quantum Rabi model in the Schrieffer-Wolff frame, while black dotted curves show the dispersive rotating-wave prediction. The local quadrature phases satisfy $\phi_a+\phi_b=\arg\!\left[\braket{a(t)b(t)}\right]+\pi$, following the instantaneous squeezed axes (see Appendix~\ref{app:squeeze}). The gray horizontal dashed line denotes the two-mode-vacuum variance, $\braket{\psi_0|\Delta X^2|\psi_0}=\braket{\psi_0|\Delta P^2|\psi_0}=1/2\equiv\braket{\Delta Q^2}_0$, and values below this level indicate squeezing. Panels (a)--(c) correspond to the stable, critical, and dynamically unstable parameter sets of Fig.~\ref{fig:paircreation}(b), respectively.}
    \label{fig:squeeze}
\end{figure*}

The stable, critical, and dynamically unstable regimes identified here are formally connected to, but distinct from, the normal-to-superradiant transition in the single-mode quantum Rabi model. In the conventional quantum Rabi model, the approximate quadratic Hamiltonian describing fluctuations about the parity-symmetric normal-phase configuration with $\braket{a}=0$ remains stable below the critical coupling, develops a soft mode at the critical point, and becomes dynamically unstable when formally continued beyond the transition~\cite{hwang2015qpt,shen2021twomoderabi}. The full Rabi Hamiltonian then stabilizes the dynamics around two parity-related displaced configurations, which constitute the superradiant phase. In the present opposite-sign two-mode model, the condition $|\lambda|=|\Delta\widetilde{\nu}|/2$ similarly marks the loss of stability of the vacuum-centered quadratic dynamics. However, the region $|\lambda|>|\Delta\widetilde{\nu}|/2$ should not by itself be identified as a superradiant phase. In this regime, the quadratic rotating-wave Hamiltonian predicts unbounded pair amplification rather than a stable symmetry-broken state. Therefore, the instability in our model is more appropriately interpreted as the dynamical instability of the normal-phase configuration.

Fig.~\ref{fig:paircreation} shows numerical simulations of spin-mediated bosonic pair dynamics in the opposite-sign two-mode quantum Rabi model of Eq.~\eqref{eq:2qrm}. The full-model numerical results are evaluated in the Schrieffer-Wolff frame to permit a direct comparison with the effective quadratic predictions; details are provided in Appendix~\ref{app:sw_frame_numerics}. The parameters are chosen in the dispersive regime of Eq.~\eqref{eq:dispersiveconditions}, with $\omega_0\gg\nu$, where $\nu$ denotes the characteristic bosonic frequency. The full-model dynamics are compared with the predictions of the effective quadratic Hamiltonian in Eq.~\eqref{eq:detuned_pair_hamiltonian}. Starting from the two-mode vacuum, the pair Hamiltonian conserves $n_a-n_b$, so the dynamics remain in the $n_a=n_b$ subspace and $\braket{n_a+n_b}=2\braket{n_a}=2\braket{n_b}$.

We compare the total mean occupation of the two modes, $\braket{n_a+n_b}$, in the all-positive-frequency and opposite-sign models in Fig.~\ref{fig:paircreation}(a). In the all-positive-frequency case, the pair-creation and pair-annihilation terms are far off resonance, while the near-resonant exchange interaction leaves the two-mode vacuum approximately unexcited. By contrast, when $\nu_a=-\nu_b$, the pair terms become near resonant and the exchange terms are off resonant. For the chosen parameters, the residual dressed pair detuning is $\Delta\widetilde{\nu}=0.04\nu$, and $\lambda=0.02\nu$. As shown in the figure, this condition of $\Delta\widetilde{\nu}=2|\lambda|$ places the system at the critical boundary, where the total mean occupation grows quadratically in time.

Moreover, Fig.~\ref{fig:paircreation}(b) compares the distinct behaviors of the stable, critical, and dynamically unstable regimes at a fixed pair coupling $\lambda=0.08\nu$. In the stable regime, $\Delta\widetilde{\nu}=0.34\nu>2|\lambda|$, we observe bounded oscillatory pair creation. The deviations from the ideal pair-Hamiltonian prediction are particularly visible for the asymmetric choice of $g_b=4g_a$, for which the spurious single-mode-squeezing strength associated with mode $b$, $g_b^2/\omega_0=0.16\nu$, is comparable to both $\lambda=0.08\nu$ and the ideal stable oscillation frequency $\Omega_{\rm st}=0.15\nu$. Residual exchange processes and higher-order dispersive corrections can also modify the dynamics. At the critical boundary, $\Delta\widetilde{\nu}=2|\lambda|=0.16\nu$, giving quadratic growth. On the other hand, the bare frequency of mode $b$ is adjusted to compensate for the leading-order dispersive shifts in the dynamically unstable regime, giving $\Delta\widetilde{\nu}=0$ and exponential pair amplification. As the phonon occupation grows, the matrix elements $g_k\sqrt{N(t)+1}$ increase, and the dispersive condition progressively weakens. Higher-order Schrieffer-Wolff corrections and residual off-resonant terms therefore become increasingly important, causing the full-model dynamics to deviate from the ideal quadratic prediction at longer times,  particularly at $\nu t \gtrsim 15$ for this unstable condition (see Appendix~\ref{app:higher_order_SW}).

In Fig.~\ref{fig:paircreation}(c), we summarize the stable and dynamically unstable regimes, separated by critical lines, in a dynamical stability diagram parameterized by the normalized bare frequency mismatch and the ratio of spin-boson coupling strengths. For the bare opposite-sign condition $\nu_a=-\nu_b$, the dressed mismatch is $\Delta\widetilde{\nu}=2s(g_a^2+g_b^2)/\omega_0$. The effective model is therefore stable for $|g_a|\neq|g_b|$ and lies at the critical boundary when $|g_a|=|g_b|$. The diagram is shown for $s=+1$; for $s=-1$, it is reflected vertically about the bare opposite-sign line $\nu_a=-\nu_b$.

It is worth noting that, within a fixed spin sector, the quadratic pair Hamiltonian evolves the two-mode vacuum into a two-mode-squeezed vacuum, up to local phase rotations, whenever the pair amplitude is nonzero. To characterize the resulting two-mode squeezing, we introduce the phase-rotated quadratures $X_k(\phi_k)$ $=[k e^{-i\phi_k}+k^\dagger e^{i\phi_k}]/\sqrt{2}$ and $P_k(\phi_k)=X_k(\phi_k+\pi/2)$ for $k=a$ and $b$ operators, where $\phi_k$ denotes the corresponding local motional phase. We then define the commuting collective quadratures $X=[X_a(\phi_a)+X_b(\phi_b)]/\sqrt{2}$ and $P=[P_a(\phi_a)-P_b(\phi_b)]/\sqrt{2}$. The local phases are chosen such that $\phi_a+\phi_b=\arg\!\left[\braket{a(t)b(t)}\right]+\pi$, which aligns the measurement axes with the instantaneous squeezed quadratures. Starting from the two-mode vacuum, the quadratic effective model produces an ideal two-mode-squeezed state with the quadrature variances $\braket{\Delta X^2}=\braket{\Delta P^2}=(1/2)e^{-2r(t)}$, where $r(t)\equiv\operatorname{arsinh}\sqrt{\braket{n_k(t)}}$ (see Appendix~\ref{app:squeeze}).

Fig.~\ref{fig:squeeze} compares the collective-quadrature variances obtained from the full two-mode quantum Rabi model of Eq.~\eqref{eq:2qrm} with the dispersive rotating-wave prediction of Eq.~\eqref{eq:detuned_pair_hamiltonian}. The two-mode vacuum variance is $\braket{\Delta Q^2}_0=1/2$, so values below this level indicate collective squeezing. In the stable regime, $\braket{n_k(t)}$ and hence $r(t)$ remain bounded and periodically return to zero, causing the collective variances to return to the vacuum level. At criticality and in the dynamically unstable regime, the growing occupation increases $r(t)$ and suppresses the collective variances below the vacuum level. The full-model results reproduce these three characteristic behaviors, with quantitative deviations from the ideal quadratic prediction due to corrections discussed in Appendix~\ref{app:higher_order_SW}.

The same collective-quadrature variances in Fig.~\ref{fig:squeeze} provide an entanglement witness through the Duan-Simon quantity~\cite{duan2000inseparability,simon2000peres},
\begin{equation}
\mathcal{V}_{\rm DS}=\braket{\Delta X^2}+\braket{\Delta P^2}.
\end{equation}
With our quadrature normalization, every separable two-mode continuous-variable state satisfies $\mathcal{V}_{\rm DS}\geq1$. For the ideal quadratic pair dynamics, $\mathcal{V}_{\rm DS}=e^{-2r(t)}$, and therefore $\mathcal{V}_{\rm DS}<1$ whenever $r(t)>0$. As shown in Fig.~\ref{fig:squeeze}, the full-model results violate the Duan-Simon bound during the time intervals in which the sum of the numerically calculated collective-quadrature variances falls below unity, thereby
certifying bipartite motional entanglement.

\subsection{Competing spin branches}
\label{sec:spincontrol}

The preceding analysis assumes that the spin is initialized in an eigenstate of $\sigma_z$, so that the bosonic dynamics is governed by a single effective Hamiltonian $H_{\rm RWA}^{(s)}$. A qualitatively different situation arises when the spin is prepared in a coherent superposition,
\begin{equation}
\ket{\psi(0)}_{\rm s}=\alpha\ket{\uparrow}+\beta\ket{\downarrow},
\qquad |\alpha|^2+|\beta|^2=1,
\label{eq:coherent_initial}
\end{equation}
while the bosonic modes are initialized in the two-mode vacuum, $\ket{0_a,0_b}$. Since the effective Hamiltonian in Eq.~\eqref{eq:detuned_pair_hamiltonian} is diagonal in the spin basis, we work in an interaction picture with respect to the branch-dependent
constant energy
\begin{equation}
H_{\rm const}
=\left[\frac{\omega_0}{2}+\frac{g_a^2+g_b^2}{\omega_0}\right]\sigma_z.
\label{eq:branchdependent_constenergy}
\end{equation}
In this interaction picture, the combined system evolves according to
\begin{equation}
H_{\rm eff}=\ket{\uparrow}\bra{\uparrow}\otimes H_{\rm RWA}^{(+1)}+\ket{\downarrow}\bra{\downarrow}\otimes H_{\rm RWA}^{(-1)},
\label{eq:hamiltonian_effective_spincontrol}
\end{equation}
giving the entangled state,
\begin{equation}
\ket{\Psi(t)}=\alpha\ket{\uparrow}\otimes\ket{\phi_{+1}(t)}+\beta\ket{\downarrow}\otimes\ket{\phi_{-1}(t)},
\label{eq:conditional_state}
\end{equation}
where $\ket{\phi_s(t)}=e^{-iH_{\rm RWA}^{(s)}t}\ket{0_a,0_b}$. We note that, strictly speaking, the initial spin state $\ket{\psi(0)}_{\rm s}$ should be defined in the dressed frame associated with the Schrieffer-Wolff transformation in Eq.~\eqref{eq:expanded_SW_generator}. In the dispersive regime considered here, however, the corrections induced by this transformation are small, so the dressed-frame initial state can be approximated by the corresponding laboratory-frame state.

The two conditional bosonic Hamiltonians $H_{\rm RWA}^{(s)}$ differ in two ways. First, the pair interaction changes sign through the factor $s\lambda$. This sign change amounts to a change of the squeezing phase (see Appendix~\ref{app:squeeze}) and, for a fixed dressed pair detuning, does not modify the bosonic occupations, since the occupation dynamics depend on $\lambda^2$. More importantly, the dispersive shifts $\widetilde{\nu}_k(s)$ and thus the residual pair detuning $\Delta\widetilde{\nu}(s)$ also depend on the spin projection, making the instability condition itself depend on the spin state. For $|\Delta\widetilde{\nu}(s)|<2|\lambda|$, the corresponding instability rate is
\begin{equation}
\Gamma_{\rm inst}(s)=\sqrt{\lambda^2-\frac{\Delta\widetilde{\nu}(s)^2}{4}},
\label{eq:spin_instability}
\end{equation}
whereas, for $|\Delta\widetilde{\nu}(s)|>2|\lambda|$, the dynamics remains bounded and oscillatory.

The two spin components can therefore belong to different dynamical regimes (see Fig.~\ref{fig:stabilitydiagram} in Appendix~\ref{app:pair_amplification}). For example, one branch may satisfy $|\Delta\widetilde{\nu}(+1)|<2|\lambda|$, while the other satisfies $|\Delta\widetilde{\nu}(-1)|>2|\lambda|$. In this case, the $s=+1$ component undergoes exponential pair amplification, while the $s=-1$ component exhibits only stable oscillatory dynamics. Likewise, one branch may lie exactly at the critical point, $|\Delta\widetilde{\nu}(+1)|=2|\lambda|$, while the other remains stable. At the symmetric point $|g_a|=|g_b|\ne 0$ and $\nu_a+\nu_b=0$, both branches are critical. A coherent spin superposition therefore realizes a quantum superposition of different instability classes, with the spin acting as a quantum control over the dynamical regime of the bosonic subsystem.

The evolution of the coherent spin superposition thus generates spin-boson entanglement. After tracing over the bosonic modes, the reduced spin density matrix becomes
\begin{equation}
\rho_{\rm s}(t)=
\begin{pmatrix}
\rho_{\uparrow\uparrow} &\rho_{\uparrow\downarrow}\\
\rho_{\downarrow\uparrow}&\rho_{\downarrow\downarrow}
\end{pmatrix}
=
\begin{pmatrix}
|\alpha|^2 &\alpha\beta^*\Phi(t)\\
\alpha^*\beta\Phi^*(t)&|\beta|^2
\end{pmatrix},
\label{eq:reduced_spin_density}
\end{equation}
where $\Phi(t)\equiv\braket{\phi_{-1}(t)|\phi_{+1}(t)}$ is the overlap of the two conditional bosonic states. As the two branches evolve into increasingly distinguishable two-mode squeezed states, the magnitude of $\Phi(t)$ decreases, leading to a loss of spin coherence through entanglement with the bosonic modes.

\begin{figure}
    \centering
    \includegraphics[width=1\linewidth]{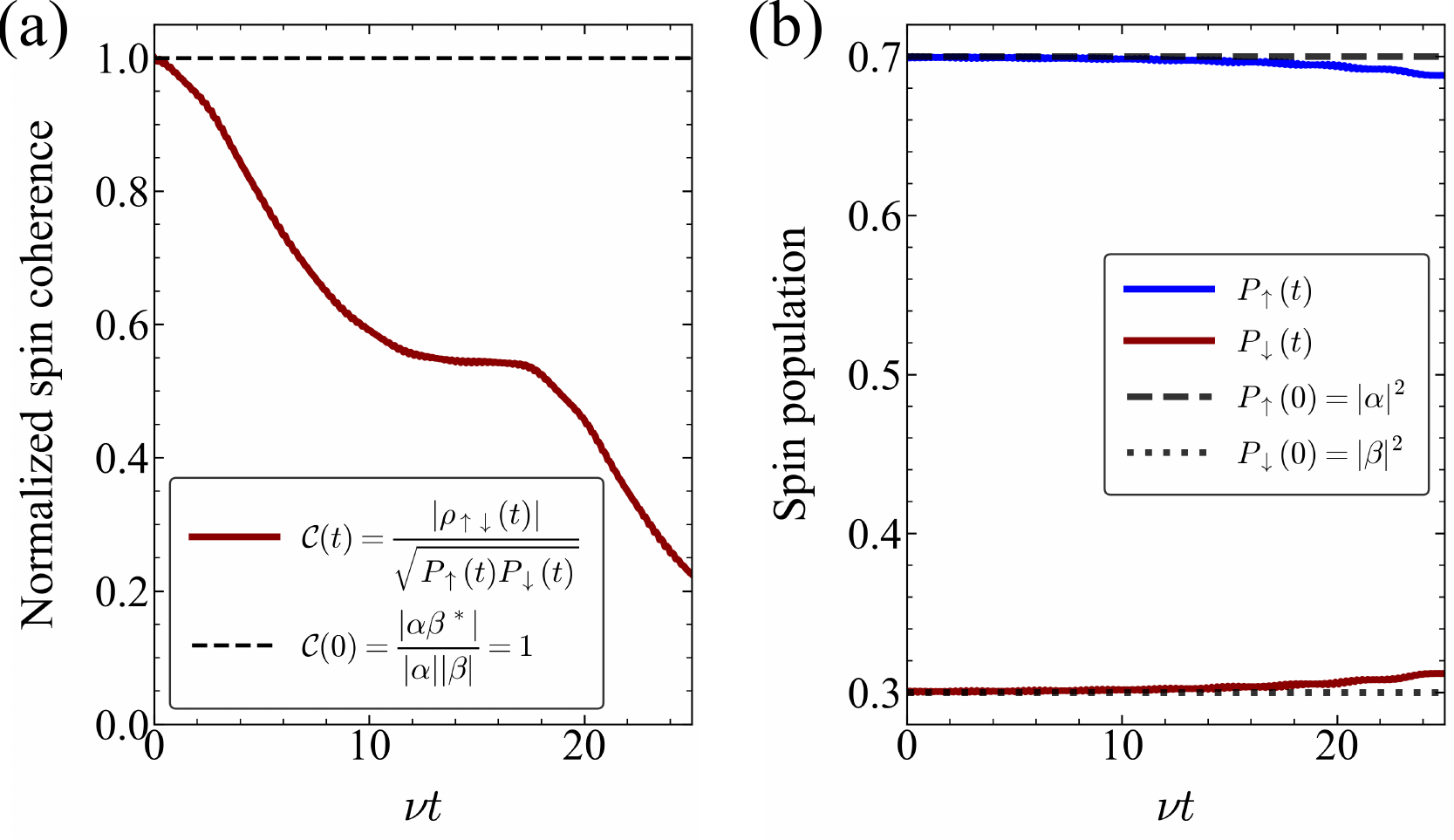}
    \caption{Spin coherence and populations of the opposite-sign two-mode quantum Rabi model. The system is initialized in $(\alpha\ket{\uparrow}+\beta\ket{\downarrow})\otimes\ket{0_a,0_b}$ with $|\alpha|^2=0.70$ and $|\beta|^2=0.30$. The parameters are $\omega_0=100\nu$, $\nu_a=\nu$, $\nu_b=-1.16\nu$, and $g_a=g_b=2\nu$, corresponding to the dispersive pair coupling $\lambda=2g_ag_b/\omega_0=0.08\nu$. (a) Population-normalized spin coherence, $\mathcal{C}(t)=|\rho_{\uparrow\downarrow}(t)|/\sqrt{P_\uparrow(t)P_\downarrow(t)}$. The dashed line marks $\mathcal{C}(0)=|\alpha\beta^*|/(|\alpha||\beta|)=1$. (b) Spin populations $P_\uparrow(t)=\rho_{\uparrow\uparrow}(t)$ and $P_\downarrow(t)=\rho_{\downarrow\downarrow}(t)$. The corresponding dashed and dotted lines denote their initial values, respectively.}
    \label{fig:spincontrol}
\end{figure}

When the spin is not measured, bosonic observables are incoherent averages over the two conditional branches. For example, the mean occupation of either mode is
\begin{equation}
\braket{n_k(t)} =|\alpha|^2\braket{n_k(t)}_{+1}+|\beta|^2\braket{n_k(t)}_{-1}.
\end{equation}
where the conditional occupations are given by the stable, critical, or unstable solutions derived in Appendix~\ref{app:pair_amplification}. By contrast, phase-sensitive Ramsey measurements of the spin, routinely available in trapped-ion experiments~\cite{monroe2021programmable}, access the real and imaginary parts of the conditional-state overlap $\Phi(t)$. For an initial equal-spin superposition, the Ramsey contrast is $|\Phi(t)|$, which, under coherent unitary evolution, determines the reduced-spin purity and thereby quantifies the spin-boson entanglement generated by the superposition of distinct spin-conditioned dynamical regimes.

To examine the competing spin branches beyond the effective quadratic approximation, we numerically evolve the two-mode quantum Rabi model from the initial state $\ket{\Psi(0)}=\left(\alpha\ket{\uparrow}+\beta\ket{\downarrow}\right)\otimes\ket{0_a,0_b}$. At any time, the full state can be decomposed exactly as
\begin{equation}
\ket{\Psi(t)}=\ket{\uparrow}\ket{\chi_{+1}(t)}+\ket{\downarrow}\ket{\chi_{-1}(t)},
\end{equation}
where the projected bosonic states are generally unnormalized. Their norms determine the instantaneous spin populations,
\begin{eqnarray}
P_\uparrow(t)&=&\braket{\chi_{+1}(t)|\chi_{+1}(t)}=\rho_{\uparrow\uparrow}(t),\\
P_\downarrow(t)&=&\braket{\chi_{-1}(t)|\chi_{-1}(t)}=\rho_{\downarrow\downarrow}(t),
\end{eqnarray}
whereas their overlap determines the off-diagonal spin coherence,
\begin{equation}
\rho_{\uparrow\downarrow}(t)=\braket{\chi_{-1}(t)|\chi_{+1}(t)}.
\end{equation}
To separate the loss of coherence caused by distinguishability of the bosonic branches from changes in the spin populations, we define the population-normalized coherence,
\begin{equation}
\mathcal{C}(t)\equiv\frac{|\rho_{\uparrow\downarrow}(t)|}{\sqrt{P_\uparrow(t)P_\downarrow(t)}}.
\label{eq:generalized_spin_coherence}
\end{equation}
This definition does not require an equal spin superposition. Positivity of the reduced density matrix guarantees $0\leq\mathcal{C}(t)\leq1$. Let us also define the normalized conditional bosonic states as
\begin{equation}
\ket{\phi_{+1}(t)}=\frac{\ket{\chi_{+1}(t)}}{\sqrt{P_\uparrow(t)}},\qquad\ket{\phi_{-1}(t)}=\frac{\ket{\chi_{-1}(t)}}{\sqrt{P_\downarrow(t)}}.
\end{equation}
We can then write $\mathcal{C}(t)=\left|\braket{\phi_{-1}(t)|\phi_{+1}(t)}\right|$. Thus, $\mathcal{C}(t)=1$ indicates identical conditional bosonic states up to a phase, while $\mathcal{C}(t)<1$ indicates that the two spin components have become correlated with distinguishable bosonic states. Since the complete state remains pure under the unitary evolution and provided both spin populations are nonzero, the mixing of the reduced spin density matrix indicated by $\mathcal{C}(t)<1$ also demonstrates spin-boson entanglement. Any subsequent revival of $\mathcal{C}(t)$ reflects a coherent partial rephasing of the two conditional bosonic branches.

In Fig.~\ref{fig:spincontrol}, we take $\omega_0=100\nu$, $\nu_a=\nu$, $g_a=g_b=2\nu$, and $\nu_b=-1.16\nu$, giving the dispersive pair coupling $\lambda=2g_ag_b/\omega_0=0.08\nu$. Within the dispersive description, these parameters give $\Delta\widetilde{\nu}(+1)=0$, so that the $s=+1$ branch lies on the unstable dressed pair resonance, whereas $\Delta\widetilde{\nu}(-1)=-0.32\nu$ places the $s=-1$ branch in the stable regime (see Fig.~\ref{fig:stabilitydiagram} in Appendix~\ref{app:pair_amplification}). Since $g_k/\omega_0=0.02$, and the appreciably populated bosonic sector remains within the dispersive regime, the corrections associated with the Schrieffer-Wolff dressing are small. We therefore perform the full-model simulation in the bare frame, so that the spin is initially in an exact coherent superposition and is uncorrelated with the bosonic modes. The initial spin populations are chosen to be $|\alpha|^2=0.70$ and $|\beta|^2=0.30$. Fig.~\ref{fig:spincontrol}(a) shows the resulting loss and partial revival of the normalized spin coherence. Since Eq.~\eqref{eq:generalized_spin_coherence} explicitly divides out the population factor $\sqrt{P_\uparrow(t)P_\downarrow(t)}$, variations in $\mathcal{C}(t)$ cannot be attributed solely to population redistribution; instead, they reflect changes in the distinguishability of the bosonic states correlated with the two spin components. A reduction below unity indicates that the conditional bosonic states have become distinguishable and, for the globally pure unitary evolution considered here, signals spin-boson entanglement. For completeness, Fig.~\ref{fig:spincontrol}(b) shows the corresponding spin populations. Their small deviations from the initial values past $\nu t \approx 15$ quantify possible spin-flip processes and other corrections beyond the dispersive regime.

\begin{figure}
    \centering
    \includegraphics[width=1\linewidth]{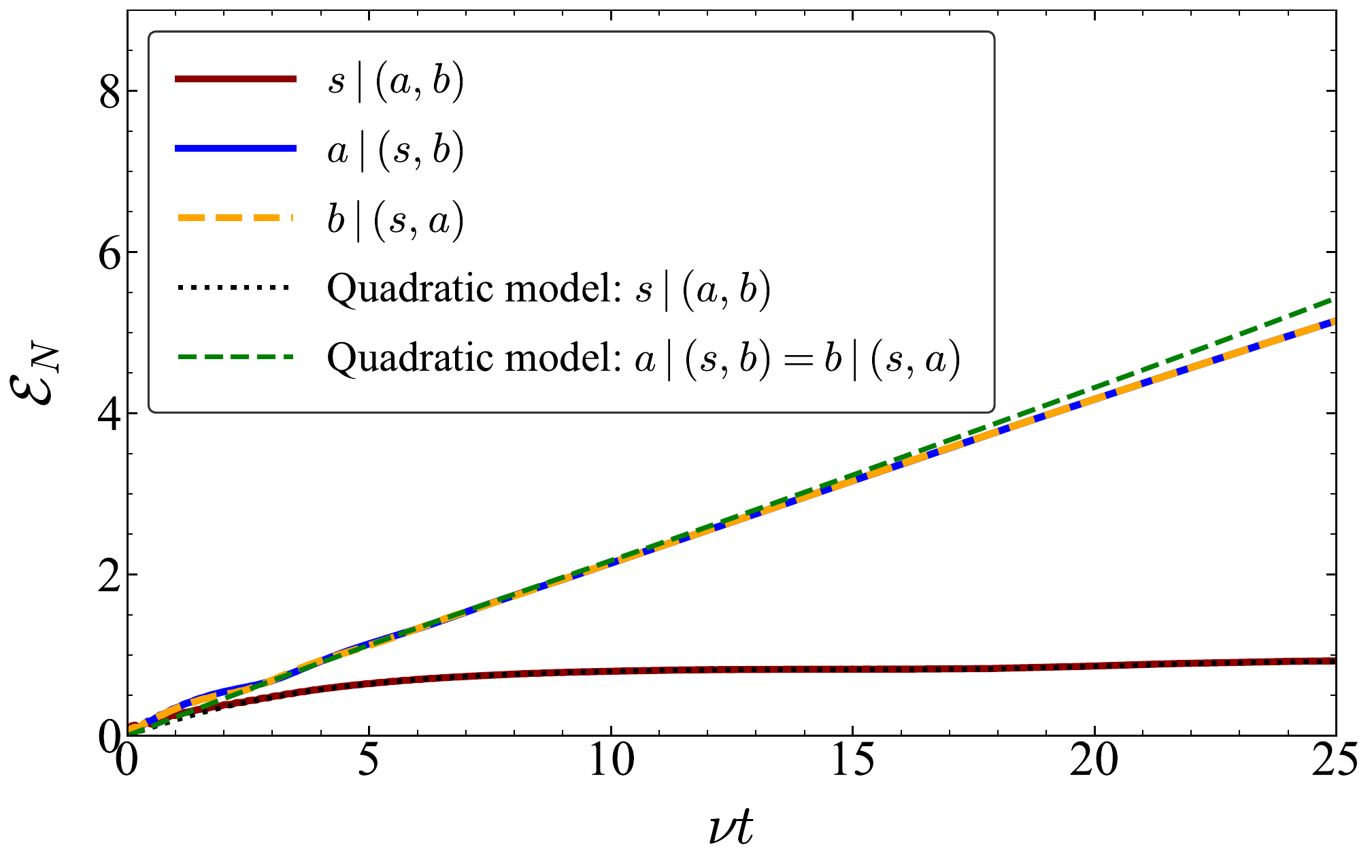}
    \caption{Spin-boson and tripartite entanglement generated by the full two-mode quantum Rabi dynamics. Logarithmic negativities are evaluated across the three bipartitions $s|(a,b)$, $a|(s,b)$, and $b|(s,a)$. The parameters and initial state are the same as in Fig.~\ref{fig:spincontrol}. Analytical predictions from the quadratic dispersive model (see Appendix~\ref{app:analytic_logneg}) are shown for comparison; within this model, $\mathcal{E}_N^{a|(s,b)}=\mathcal{E}_N^{b|(s,a)}$.}
    \label{fig:spincontrol_logneg}
\end{figure}

To verify directly that the reduction of spin coherence originates from entanglement with the bosonic modes, we calculate the logarithmic negativity~\cite{vidal2002computable,plenio2005logarithmic} for the bipartition between the spin and the two-mode bosonic subsystem,
\begin{equation}
\mathcal{E}_{N}^{s|(a,b)}(t)=\log_2\left\|\rho^{T_{\rm s}}(t)\right\|_1,
\label{eq:spin_boson_logneg}
\end{equation}
where $T_{\rm s}$ denotes the partial transpose with respect to the spin~\cite{peres1996separability}. Since the full two-mode quantum Rabi evolution is unitary, and the global state remains pure, this quantity can be evaluated directly from the reduced spin density matrix as~\cite{vidal2002computable}
\begin{equation}
\mathcal{E}_{N}^{s|(a,b)}(t)=\log_2\!\left[1+2\sqrt{P_\uparrow(t)P_\downarrow(t)-\left|\rho_{\uparrow\downarrow}(t)\right|^2}\right].
\label{eq:spin_boson_logneg_reduced}
\end{equation}
The derivation from Eq.~\eqref{eq:spin_boson_logneg} to Eq.~\eqref{eq:spin_boson_logneg_reduced} is available in Appendix~\ref{sec:lognegder}. Equivalently, by using the normalized coherence introduced in Eq.~\eqref{eq:generalized_spin_coherence}, we get
\begin{equation}
\mathcal{E}_{N}^{s|(a,b)}(t)=\log_2\!\left[1+2\sqrt{P_\uparrow(t)P_\downarrow(t)\left(1-\mathcal{C}^2(t)\right)}\right].
\label{eq:exp_logneg_sab}
\end{equation}
Here, $\mathcal{E}_{N}^{s|(a,b)}$ can be evaluated from the spin populations and coherence alone, without tomography of the two motional modes. Thus, the spin efficiently acts both as a quantum control of the bosonic stability regime and as a local probe of the spin-boson entanglement generated by the competing branch-dependent dynamics. The nonzero logarithmic negativity shown in Fig.~\ref{fig:spincontrol_logneg} directly confirms that the loss of normalized spin coherence in Fig.~\ref{fig:spincontrol} is accompanied by entanglement generation between the spin and the bosonic modes. Its nonmonotonic behavior reflects the coherent separation and partial re-overlap of the bosonic states correlated with the two spin components. We note that Eq.~\eqref{eq:spin_boson_logneg_reduced} is exact only under global purity. With uncontrolled decoherence in the system, the reduced spin state no longer uniquely determines the spin-boson entanglement and instead provides an inferred estimate of $\mathcal{E}_{N}^{s|(a,b)}$.   

Since the pair dynamics simultaneously generate strong correlations between the two bosonic modes, the evolved state can develop genuine tripartite entanglement among the spin and the two motional modes. To establish this directly, we additionally calculate the logarithmic negativities across the remaining two bipartitions,
\begin{align}
\mathcal{E}_{N}^{a|(s,b)}=\log_2\left\|\rho^{T_a}\right\|_1,\quad
\mathcal{E}_{N}^{b|(s,a)}=\log_2\left\|\rho^{T_b}\right\|_1.
\end{align}
As shown in Fig.~\ref{fig:spincontrol_logneg}, the initially separable state has vanishing logarithmic negativity across all three bipartitions. During the evolution, they become simultaneously nonzero. Since the global spin-boson state remains pure under the unitary dynamics considered here, this demonstrates entanglement across each of the three bipartitions, $s|(a,b)$, $a|(s,b)$, and $b|(s,a)$, and therefore establishes genuine tripartite entanglement among the spin and the two bosonic modes~\cite{guhne2009entanglement,huber2010genuine}. The different magnitudes of the three logarithmic negativities are also expected from their Hilbert-space dimensions. Since the spin is a two-level system, the Schmidt rank across the $s|(a,b)$ bipartition is at most two, which bounds $\mathcal{E}_{N}^{s|(a,b)}\leq\log_2 2=1$~\cite{ekert1995entangled,vidal2002computable}. By contrast, the bosonic modes have much larger Hilbert spaces, so $\mathcal{E}_{N}^{a|(s,b)}$ and $\mathcal{E}_{N}^{b|(s,a)}$ are not subject to this qubit bound and can grow substantially as the pair occupation increases. The nearly identical values of $\mathcal{E}_N^{a|(s,b)}$ and $\mathcal{E}_N^{b|(s,a)}$ also reflect the correlated pair dynamics of the two motional modes.

The spin-conditioned dynamics also provide a route to preparing non-Gaussian entangled states of the two bosonic modes. Within the quadratic description, each conditional state $\ket{\phi_s(t)}$ is Gaussian because the quadratic Hamiltonian $H_{\rm RWA}^{(s)}$ generates a Gaussian unitary acting on the initial two-mode vacuum~\cite{caves1985twophoton1,schumaker1985twophoton2,weedbrook2012gaussian}. However, measuring the spin in a basis different from the $\sigma_z$ basis coherently combines the two spin-conditioned bosonic branches. For the spin-analysis basis $\ket{\pm_\theta}=[\ket{\uparrow}\pm e^{i\theta}\ket{\downarrow}]/\sqrt{2}$, the conditional bosonic state is
\begin{equation}
\ket{\psi_{\pm,\theta}(t)}=\frac{\alpha\ket{\phi_{+1}(t)}\pm e^{-i\theta}\beta\ket{\phi_{-1}(t)}}{\sqrt{1\pm2\,{\rm Re}\!\left[e^{i\theta}\alpha\beta^*\Phi(t)\right]}}.
\label{eq:postselected_bosonic_state_general}
\end{equation}

As shown in Appendix~\ref{app:analytic_logneg}, we can write each conditional Gaussian state as a two-mode-squeezed vacuum,
\begin{equation}
\ket{\phi_s(t)} = e^{i\vartheta_s(t)}\sqrt{1-|\zeta_s(t)|^2}\sum_{n=0}^{\infty}\zeta_s^n(t)\ket{n,n},
\label{eq:main_conditional_tmsv}
\end{equation}
where the complex parameter $\zeta_s(t)$ determines the squeezing magnitude and phase of spin branch $s$, and $\vartheta_s(t)$ is the overall phase accumulated during the evolution. For simplicity, we consider postselection in the $\sigma_x$ basis, corresponding to $\theta=0$. The postselected state then becomes
\begin{equation}
\ket{\psi_\pm(t)}=\sum_{n=0}^{\infty}c_n^{(\pm)}(t)\ket{n,n},
\end{equation}
with
\begin{equation}
c_n^{(\pm)}(t)=\frac{A(t)\zeta_{+1}^n(t)\pm B(t)\zeta_{-1}^n(t)}{\sqrt{1\pm2\,{\rm Re}\!\left[\alpha\beta^*\Phi(t)\right]}},
\label{eq:postselected_cn}
\end{equation}
where
\begin{equation}
\begin{aligned}
A(t) &= \alpha e^{i\vartheta_{+1}(t)}
\sqrt{1-|\zeta_{+1}(t)|^2}, \\
B(t) &= \beta e^{i\vartheta_{-1}(t)}
\sqrt{1-|\zeta_{-1}(t)|^2}.
\end{aligned}
\end{equation}
Thus, when the two spin branches acquire different squeezing magnitudes or phases, the postselected Fock-state amplitudes are a coherent sum of two distinct geometric sequences, and the resulting bosonic state is generally non-Gaussian. In particular, the correlated Fock-state amplitudes of a two-mode-squeezed vacuum form a single geometric sequence and therefore satisfy $c_0c_2=c_1^2$~\cite{caves1985twophoton1,schumaker1985twophoton2,
botero2003modewise}. By contrast, Eq.~\eqref{eq:postselected_cn} gives
\begin{equation}
c_0^{(\pm)}c_2^{(\pm)}-\left[c_1^{(\pm)}\right]^2=\frac{\pm A(t)B(t)\left[\zeta_{+1}(t)-\zeta_{-1}(t)\right]^2}{1\pm2\,{\rm Re}\!\left[\alpha\beta^*\Phi(t)\right]}.
\label{eq:nongaussian_condition}
\end{equation}
For a nonzero postselection probability, this quantity is nonzero whenever both branches contribute and have distinct squeezing parameters, $\zeta_{+1}(t)\neq\zeta_{-1}(t)$. The postselected state is therefore non-Gaussian. Spin measurement, therefore, converts the quantum control over the bosonic stability dynamics into a measurement-induced resource for preparing non-Gaussian two-mode states.

Another important consequence is that the entanglement of these postselected states is no longer fully characterized by their first- and second-order quadrature moments~\cite{shchukin2005inseparability, hertz2016detection}. While the Duan-Simon criterion is necessary and sufficient for separability of the symmetric two-mode Gaussian states~\cite{duan2000inseparability,simon2000peres}, for non-Gaussian states, $\mathcal{V}_{\rm DS}\geq1$ does not imply separability~\cite{hertz2016detection}. Thus, the postselected dynamics can generate bosonic entanglement that is not detected by the Duan-Simon criterion.

As an illustrative example, consider two conditional Gaussian states with squeezing parameters $\zeta_{+1}=\zeta$ and $\zeta_{-1}=-\zeta$, where $0<|\zeta|<1$. This is an idealized construction and not an exact output of the time-independent dispersive rotating-wave dynamics given by Eq.~\eqref{eq:detuned_pair_hamiltonian}. For equal spin amplitudes, $|\alpha|=|\beta|=1/\sqrt{2}$, the spin-analysis phase $\theta$ can be chosen so that the two contributions to the $+$ postselection outcome have the same overall phase. The two branches then interfere constructively for even pair-number components and destructively for odd ones, yielding the non-Gaussian state
\begin{equation}
\ket{\psi_{\rm even}}=\sqrt{1-|\zeta|^4}\sum_{m=0}^{\infty}\zeta^{2m}\ket{2m,2m}.
\label{eq:main_even_pair_state}
\end{equation}
For any $0<|\zeta|<1$, more than one correlated Fock-state component is present, and the two bosonic modes are therefore entangled. However, the absence of odd pair-number components causes the anomalous second-order correlation $\langle ab\rangle$ to vanish. As derived in Appendix~\ref{app:postselected_non_gaussian}, the minimum Duan-Simon quantity is consequently
\begin{equation}
\mathcal{V}_{\rm DS,min}^{(\rm even)}=\frac{1+3|\zeta|^4}{1-|\zeta|^4}>1,
\label{eq:main_even_pair_DS}
\end{equation}
while the logarithmic negativity remains finite,
\begin{equation}
E_{\mathcal N}^{a|b,(\rm even)}=\log_2\left(\frac{1+|\zeta|^2}{1-|\zeta|^2}\right)>0.
\label{eq:main_even_pair_logneg}
\end{equation}
Thus, the postselected bosonic state is entangled while never violating the Duan-Simon inequality.

\begin{figure}
    \centering
    \includegraphics[width=1\linewidth]{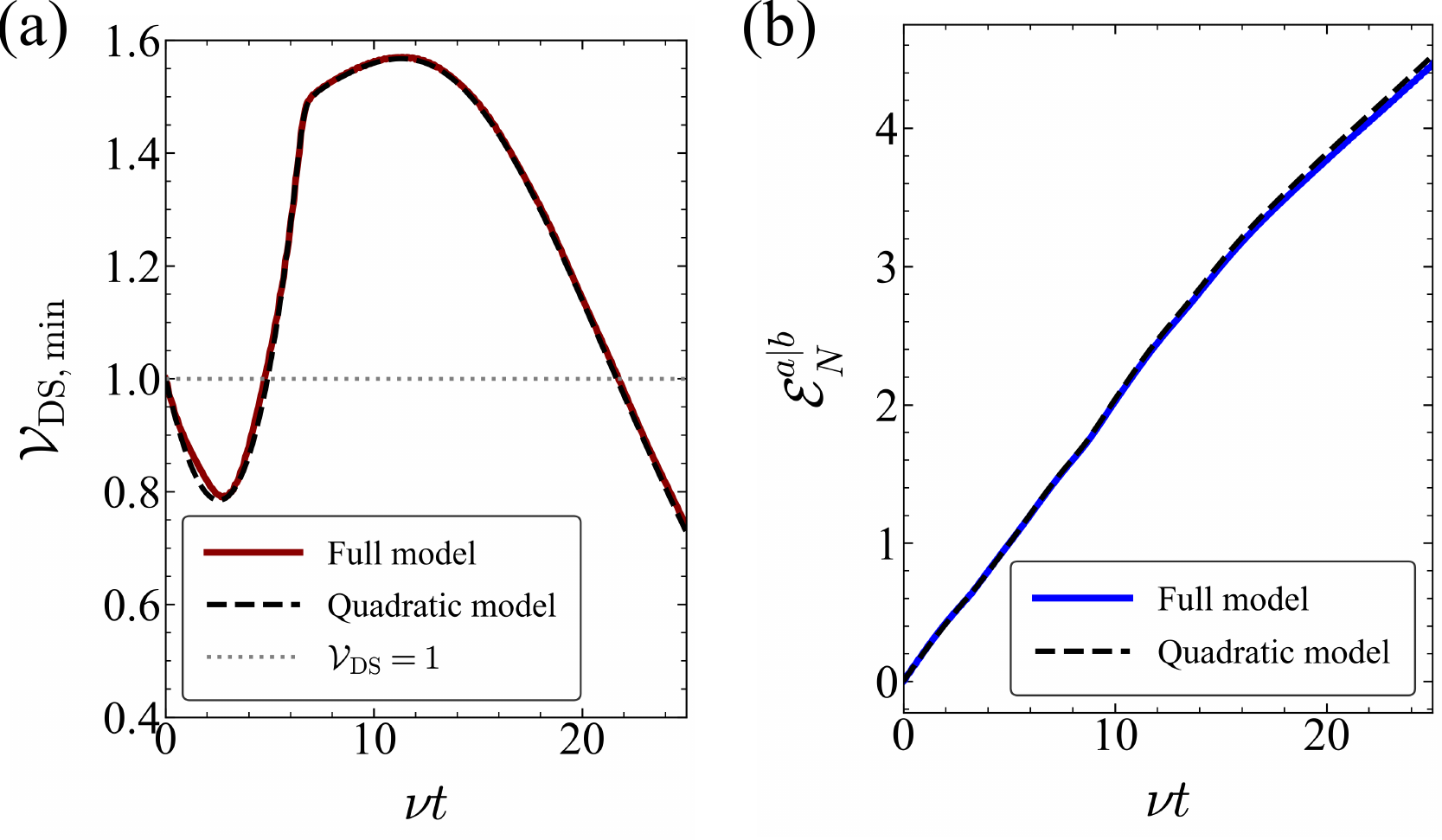}
    \caption{Spin-postselected bosonic entanglement. (a) Minimum Duan-Simon quantity $\mathcal{V}_{\rm DS,min}$ and (b) logarithmic negativity $\mathcal{E}_{N}^{a|b}$ of the two bosonic modes following postselection of the spin onto the $+$ outcome. The parameters are $\omega_0=100\nu$, $g_a=g_b=2\nu$, $\nu_a=\nu$, and $\nu_b=-1.04\nu$. Solid curves show the full two-mode quantum Rabi model, while dashed curves represent the quadratic dispersive model. The spin is initialized with equal amplitudes, $\alpha=\beta=1/\sqrt{2}$, and the spin-analysis phase is chosen as $\theta(t)=\pi/2+\omega_{\rm spin,eff}t$, where $\omega_{\rm spin,eff}=\omega_0+2(g_a^2+g_b^2)/\omega_0$ accounts for the leading-order relative spin precession. The dotted gray line in (a) marks the Duan-Simon threshold $\mathcal{V}_{\rm DS}=1$.}
    \label{fig:postselected_non_gaussian}
\end{figure}

We next examine whether this behavior also naturally occurs in the dynamically generated postselected states, without imposing $\zeta_{-1}=-\zeta_{+1}$. In Fig.~\ref{fig:postselected_non_gaussian}, we compare the postselected bosonic dynamics of the full two-mode quantum Rabi model with those of the quadratic dispersive model. We choose $\omega_0=100\nu$, $g_a=g_b=2\nu$, $\nu_a=\nu$, and $\nu_b=-1.04\nu$. For these parameters, the $s=+1$ branch lies in the unstable regime, while the $s=-1$ branch remains stable. The spin is initialized with equal amplitudes, $\alpha=\beta=1/\sqrt{2}$, and, at each evolution time, the spin-analysis phase is chosen as
\begin{equation}
\begin{aligned}
    \theta(t)&= \theta_0+\frac{\pi}{2}+\omega_{\mathrm{spin,eff}}t,\\
    \omega_{\mathrm{spin,eff}}&= \omega_0+\frac{2(g_a^2+g_b^2)}{\omega_0},
\end{aligned}
\end{equation}
where $\theta_0=\arg(\beta)-\arg(\alpha)=0$ for the initial state. This choice tracks the dispersively shifted spin precession generated by Eq.~\eqref{eq:branchdependent_constenergy}, giving the fixed interaction-picture analysis phase $\theta_{\mathrm{I}}(t)=\theta(t)-\omega_{\mathrm{spin,eff}}t=\theta_0+\pi/2$. The $\pi/2$ offset sets the relative phase imposed by the spin measurement. For the $+$ outcome, we evaluate the resulting state's minimum Duan-Simon quantity $\mathcal{V}_{\rm DS,min}$ and logarithmic negativity $\mathcal{E}_{N}^{a|b}$.

The full two-mode quantum Rabi results closely follow the quadratic predictions for both observables throughout the simulated time window. In both descriptions, $\mathcal{V}_{\rm DS,min}$ exceeds unity over an intermediate interval while $\mathcal{E}_{N}^{a|b}$ remains positive and continues to grow. Thus, the non-Gaussian bosonic states generated by spin postselection can exhibit increasing entanglement without violating the Duan-Simon inequality, motivating entanglement tests beyond second moments~\cite{shchukin2005inseparability,hertz2016detection}.

\subsection{Experimental design}
\label{sec:experiment}

An experimental demonstration of pair creation across the stable, critical, and dynamically unstable regimes discussed in Sec.~\ref{sec:paircreation}, as well as the competing spin-conditional dynamics described in Sec.~\ref{sec:spincontrol}, should be feasible with current trapped-ion platforms. Table~\ref{table:exprange} summarizes representative ranges of the relevant parameters accessible in analog trapped-ion simulators~\cite{padilla2025delocalizedexcitationtransferopen}.

\begin{table}[h!]
\centering
\begin{tabular}{lll}
\toprule
\textbf{System parameter} & \thickspace \textbf{Experimental range} \\
\text{($\hbar = 1$)} & \thickspace \text{$[2\pi \times ~\text{kHz}]$}\\
\midrule
Spin-energy splitting $\omega_0$ & \thickspace $\le 500$~\footnote{Since the Hamiltonian-engineering requirements for the proposed single-spin implementation are less restrictive than those for the multispin systems of Ref.~\cite{padilla2025delocalizedexcitationtransferopen}, larger carrier Rabi frequencies or common frequency offsets may be used, allowing spin-energy splittings beyond the range considered there.} \\
Effective mode frequency $\nu_k$ & \thickspace $2\text{--}40$\\ 
Spin-boson coupling $g_k$ & \thickspace $\lesssim 40$\\
\bottomrule
\end{tabular}
\caption{Representative ranges for the relevant system parameters accessible by analog trapped-ion quantum simulators. The ranges for $\nu_k$ and $g_k$ are based on Ref.~\cite{padilla2025delocalizedexcitationtransferopen}; see footnote for the quoted range of $\omega_0$.}
\label{table:exprange}
\end{table}

One possible implementation consists of two ions, one serving as the qubit ion and the other as an ancilla used to characterize the shared motional modes. The two-mode quantum Rabi model can be implemented using two independently controlled Raman spin-dependent forces on the qubit ion, each addressing a motional mode. Their detunings and amplitudes set the signed effective frequencies and spin-boson couplings, respectively. The effective spin-energy splitting can be generated either by a carrier drive or by a common frequency offset, as described in Sec.~\ref{sec:hamiltonianengineering}. Following motional ground-state cooling and standard qubit initialization by optical pumping~\cite{leibfried2003ion}, the Schrieffer-Wolff-frame dynamics could be accessed experimentally using the following sequence,
\begin{align}
U_{\rm seq}(t)&=e^S R e^{-iH_{\rm exp}t} R^\dagger e^{-S}\nonumber\\
&=e^S e^{-iH_{\rm 2QRM}t} e^{-S} \nonumber\\&\approx e^{-iH_{\rm eff}t},
\end{align}
where $H_{\rm exp}$ denotes the Hamiltonian implemented in the experimental spin basis, and $R$ is a carrier $\pi/2$ rotation that maps this basis onto that used for defining $H_{\rm 2QRM}$. The basis rotation is required when the effective spin-energy splitting is generated using the carrier-drive approach. For example, choosing $R=R_x\left(\pi/2\right)=e^{-i\pi\sigma_x/4}$ maps the $y$ spin axis onto the $z$ axis, $R\sigma_y R^\dagger=\sigma_z$. On the other hand, the common-frequency-offset approach does not require any basis rotation, and thus $R=\identity$.

The transformations $e^{\pm S}$ can be synthesized using phase- and amplitude-controlled bichromatic sideband drives that implement the required linear combination of spin-dependent forces along orthogonal spin and motional quadratures. Explicitly, with $x_k=X_k(0)$ and $p_k=P_k(0)$,
\begin{equation}
S\simeq i\sqrt{2}\sum_k\frac{g_k}{\omega_0}\left(\sigma_y x_k+\frac{\nu_k}{\omega_0}\sigma_x p_k\right),
\end{equation}
which is the leading-order form of Eq.~\eqref{eq:expanded_SW_generator}. The dominant contribution is a $\sigma_y$ force coupled to $x_k$, while the momentum-coupled $\sigma_x$ correction is smaller by $|\nu_k|/\omega_0$. Importantly, $S$ is defined in the same rotating frame as $H_{\rm 2QRM}$. Therefore, when $e^{\pm S}$ is implemented using standard spin-dependent-force pulses, the spin and motional phases of these drives must be referenced to the simulation frame at their respective application times. For a mode with effective frequency $\nu_k$, the relative phase between the simulation-frame motional quadratures and the phase reference of the physical sideband drive changes by $\nu_k t$ during an $H_{\rm 2QRM}$ evolution of duration $t$. Accordingly, the final $e^{S}$ pulse requires a compensating phase shift so that it implements the same simulation-frame motional quadratures as the initial $e^{-S}$ pulse. The sign of $\nu_k$ must also be retained in this phase tracking, so that positive- and negative-frequency modes accumulate relative phase in opposite directions.

After applying the inverse frame transformation $e^{S}$ at the end of the evolution, standard resolved-sideband measurements of the bare phonon occupations~\cite{leibfried2003ion} and joint characteristic-function measurements of the motional quadratures~\cite{fluhmann2020direct,jia2022multimode,whitlow2023conical,valahu2023conical,jeon2025jointparity} directly probe the corresponding observables in the effective dispersive frame. The shared-mode occupations may be read out using either ion, while the ancilla may be preferred for the characteristic-function measurements as it can be initialized to provide higher signal contrast~\cite{so2026thesis}. For the parameters considered here, the difference between the bare- and dressed-frame results is perturbatively small. Implementing $e^{\pm S}$ is therefore unnecessary for observing the qualitative dynamics, although doing so enables a more precise comparison with the effective dispersive theory.

\begin{figure}
    \centering
    \includegraphics[width=1\linewidth]{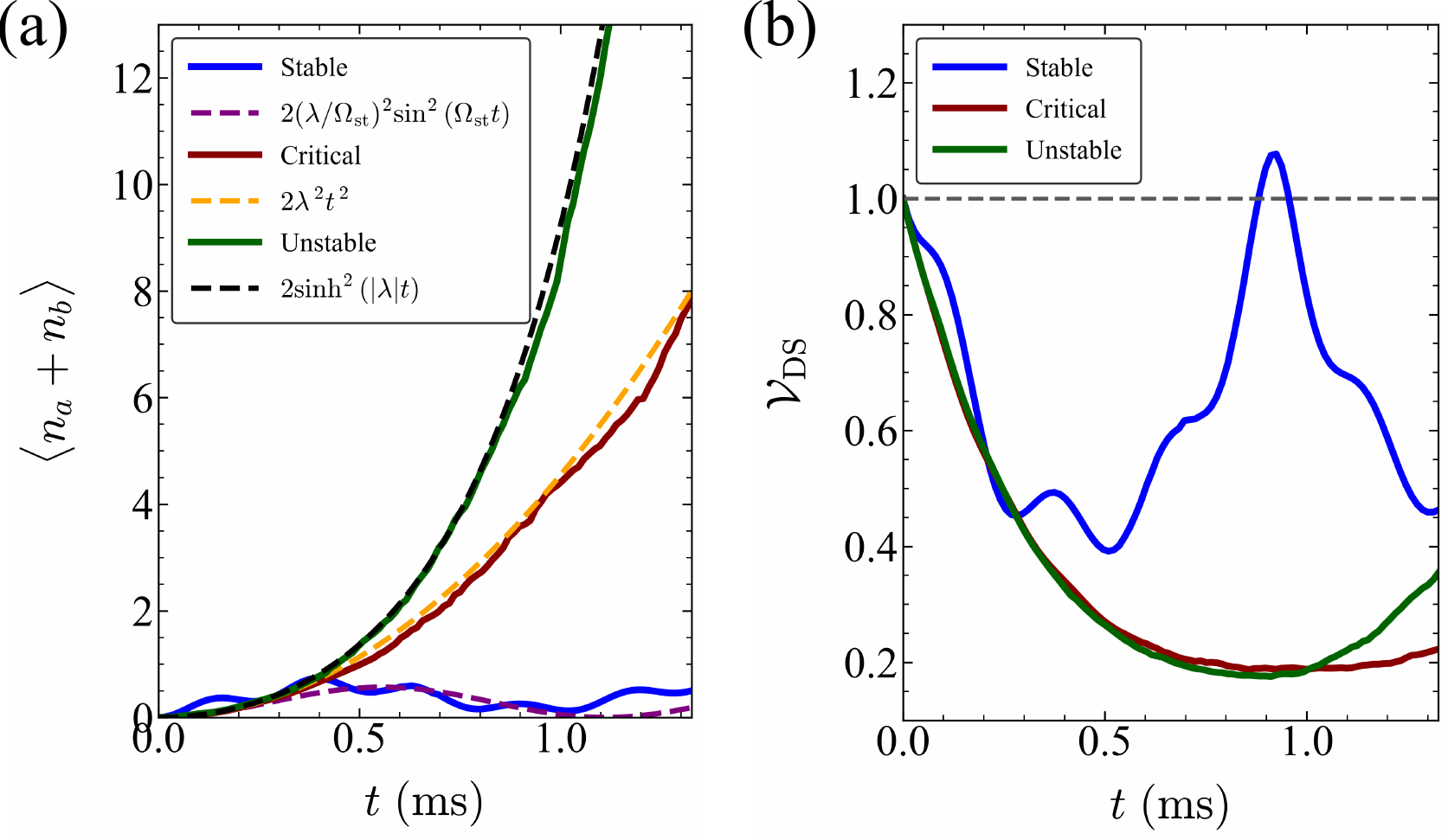}
    \caption{Proposed experiment for the opposite-sign two-mode quantum Rabi model. (a) Bosonic pair-creation and (b) Duan-Simon dynamics in the presence of motional heating and dephasing for the stable, critical, and dynamically unstable regimes. The parameters and initial state are the same as in Figs.~\ref{fig:paircreation}(b) and~\ref{fig:squeeze}, with $\nu=2\pi\times3~\mathrm{kHz}$. We include independent diffusive motional-heating channels, $\Gamma_h[\mathcal{D}_k+\mathcal{D}_{k^\dagger}]$, and motional dephasing channels, $\Gamma_\phi\mathcal{D}_{k^\dagger k}$, for each mode $k=a,b$, with $\Gamma_h=\Gamma_\phi=0.05~\mathrm{ms}^{-1}$. For these rates, the corresponding motional Ramsey coherence time is approximately $T_2\simeq(2\Gamma_h+\Gamma_\phi/2)^{-1}=8~\mathrm{ms}$~\cite{turchette2000decoherence,menon2026monolithic}, providing a conservative estimate relative to the performance achievable in current trapped-ion platforms. In panel (b), $\mathcal{V}_{\rm DS}<1$ certifies bipartite motional entanglement between the two modes. The simulations use a bosonic cutoff of $N_{\rm cut}=200$ Fock states per mode and are averaged over 200 quantum trajectories.}
    \label{fig:experimentheating}
\end{figure}

The parameter ranges listed in Table~\ref{table:exprange} indicate that, for a characteristic mode frequency of $\nu=2\pi\times 3$ kHz, the parameters required for all regimes considered in this work fall within representative ranges demonstrated in current trapped-ion platforms. Explicitly, choosing $\nu=2\pi\times3~\mathrm{kHz}$ maps the largest values used in our simulations to $\omega_0=2\pi\times300~\mathrm{kHz}$ and $g_k\leq2\pi\times12~\mathrm{kHz}$, both of which lie within the ranges listed in Table~\ref{table:exprange}. Given that state-of-the-art trapped-ion platforms have demonstrated motional coherence times of tens of milliseconds or longer~\cite{menon2026monolithic}, evolution up to $\nu t=25$ should not pose a significant limitation. For our proposed characteristic frequency, this corresponds to an evolution time of $t\approx1.3~\mathrm{ms}$, well below the reported coherence times.

We further examine numerically the effects of motional heating and dephasing, which are common decoherence mechanisms in trapped-ion spin-boson experiments~\cite{wineland1997issue,myatt2000decoherence}. More generally, in the presence of irreversible processes arising from uncontrolled environmental noise or deliberately engineered reservoirs, the open-system dynamics generated by a Hamiltonian $H$ may be described by the Lindblad master equation
\begin{align}
\frac{\partial \rho}{\partial t}&=-i[H,\rho]+\sum_c \Gamma_c \mathcal{D}_c[\rho],\label{eq:master}\\
\mathcal{D}_c[\rho]&=c\rho c^\dagger-\frac{1}{2}\left\{c^\dagger c,\rho\right\},
\end{align}
where $\Gamma_c$ denotes the rate associated with the jump operator $c$~\cite{gorini1976completely,lindblad1976generators}. In Fig.~\ref{fig:experimentheating}, we include independent diffusive motional-heating channels for the two modes,
\begin{equation}
\sum_{c_1}\Gamma_{c_1}\mathcal{D}_{c_1}[\rho]=\Gamma_h\sum_{k=a,b}\left(\mathcal{D}_k[\rho]+\mathcal{D}_{k^\dagger}[\rho]\right),
\end{equation}
with $\Gamma_h=0.05~\mathrm{ms}^{-1}$, as well as independent motional dephasing channels at the same rate ($\Gamma_\phi=0.05~\mathrm{ms}^{-1}$),
\begin{equation}
\sum_{c_2}\Gamma_{c_2}\mathcal{D}_{c_2}[\rho]=\Gamma_\phi\sum_{k=a,b}\mathcal{D}_{k^\dagger k}[\rho].
\end{equation}
These dissipative processes, $\sum_c \Gamma_c \mathcal{D}_c[\rho]=\sum_{c_1}\Gamma_{c_1}\mathcal{D}_{c_1}[\rho]+\sum_{c_2}\Gamma_{c_2}\mathcal{D}_{c_2}[\rho]$, produce only small quantitative changes in both the total bosonic occupation and the Duan-Simon quantity in the stable, critical, and dynamically unstable regimes over the simulated time window, relative to the dissipation-free results in Figs.~\ref{fig:paircreation}(b) and \ref{fig:squeeze}. In the absence of coherent dynamics, the heating channels alone increase the total two-mode occupation at a rate $2\Gamma_h=0.10~\mathrm{ms}^{-1}$, corresponding to an increase of only $\Delta\langle n_a+n_b\rangle\approx0.13$ over $t\approx1.3~\mathrm{ms}$. Fig.~\ref{fig:experimentheating} shows that key signatures, including pair-creation dynamics across different stability regimes (Fig.~\ref{fig:experimentheating}(a)) and the corresponding bipartite motional entanglement revealed through joint-mode squeezing (Fig.~\ref{fig:experimentheating}(b)), remain clearly visible even in the presence of deliberately conservative levels of experimental decoherence.

To investigate the spin-conditioned stability dynamics discussed in
Sec.~\ref{sec:spincontrol}, an arbitrary coherent spin superposition can be
prepared after motional ground-state cooling and optical pumping by applying a phase-controlled resonant carrier rotation~\cite{olmschenk2007manipulation,
wang2021single}. After the spin-boson evolution, the reduced spin density
matrix can be reconstructed using standard qubit tomography. Direct state-dependent fluorescence yields the populations $P_\uparrow$ and $P_\downarrow$, while $\pi/2$ carrier analysis pulses with phases differing by $\pi/2$ map the $\sigma_x$ and $\sigma_y$ coherences onto measurable populations~\cite{wang2021single}. These measurements determine $\rho_{\rm s}=\left(\identity+\braket{\sigma_x}\sigma_x+\braket{\sigma_y}\sigma_y+\braket{\sigma_z}\sigma_z\right)/2$, and hence provide access to both the spin populations and the coherence between the two spin-conditioned dynamical branches, which together evaluate $\mathcal{E}_{N}^{s|(a,b)}$ through Eq.~\eqref{eq:spin_boson_logneg_reduced}.

\section{Outlook}

In this work, we establish the relative signs of effective bosonic frequencies as an independently controllable parameter in trapped-ion simulations of multimode spin-boson systems. In a two-mode quantum Rabi model, reversing the sign of one mode changes the resonant spin-mediated interaction from excitation exchange to pair creation and gives rise to a dynamical instability whose threshold depends on the spin-conditioned dispersive shifts. The resulting dynamics generate two-mode squeezing and motional entanglement. In addition, coherent spin superpositions enable different spin branches to occupy distinct stability regimes. Spin postselection further converts these conditional dynamics into non-Gaussian entangled bosonic states whose entanglement need not be detected by second-moment criteria. These effects can be accessed with parameters and controls available in current trapped-ion experiments.

Beyond the dispersive regime considered here, trapped-ion implementations can access regimes in which $\omega_0$, $|\nu_k|$, and $g_k$ are all comparable, including the ultrastrong- and deep-strong-coupling regimes of the quantum Rabi model~\cite{pedernales2015quantumrabi,puebla2017probing,lv2018rabi,cai2021rabi}. In this nonperturbative regime, the spin can no longer be adiabatically eliminated, and the effective quadratic bosonic description generally breaks down. The resulting dynamics therefore provide an opportunity to determine how spin-boson entanglement, spin backaction, and higher-order processes reshape the ideal pair instability.

The same construction can be extended to larger numbers of vibrational modes and spins, as well as to multilevel internal-state manifolds, leading to multimode Rabi, Dicke-type, Rabi-Hubbard, and more general linear vibronic-coupling models containing both positive- and negative-signature bosonic sectors. Existing experiments~\cite{safavinaini2018verification,mei2022rabihubbard,wang2024simulating,so2025multimode,sun2025quantum} on Dicke, Rabi-Hubbard, and structured multimode spin-boson models demonstrate that the required multimode and multispin control is experimentally available. Introducing mixed-sign frequencies enables the study of competing unstable channels, collective dynamical behaviors, mode-selective amplification, mirrored transfer resonances, and the growth and redistribution of multipartite spin-boson entanglement. Extensions to collective-spin systems may additionally connect mixed-sign mode engineering to resonant spin-boson protocols for generating spin squeezing, where collective-spin fluctuations effectively participate in two-mode-squeezing dynamics~\cite{barberena2024fast}.

Another promising direction is to incorporate mixed-sign effective frequencies into generalized spin-dependent-force engineering. Two non-commuting linear spin-dependent forces with suitable detunings can generate effective nonlinear bosonic interactions~\cite{sutherland2021universal}, recently demonstrated experimentally up to fourth order~\cite{bazavan2026squeezing}. In a multimode setting, independently chosen frequency signs could satisfy higher-order resonance conditions, such as $p\nu_a+q\nu_b\simeq0$, generalizing the pair resonance $\nu_a+\nu_b\simeq0$. This could enable resonant multiphonon conversion, higher-order pair production, and potentially non-Gaussian dynamical instabilities.

Lastly, combining mixed-sign modes with engineered reservoirs would provide access to open-system stability boundaries and nonequilibrium steady states arising from the competition between coherent and dissipative processes~\cite{lemmer2018simulating,wang2024simulating,so2024electrontransfer,sun2025quantum,zhu2025dissipation,so2026finitetemp,kienzler2015reservoir}, as well as dissipative phase transitions and reservoir-engineered modifications of phase boundaries~\cite{hwang2018dissipative,debecker2024controlling,
debecker2024spectral,zhu2026dissipative}. In particular, introducing dissipation would allow one to study how the closed-system instability generated here by spin-mediated coupling between mixed-sign bosonic modes is modified by losses, bringing the trapped-ion simulation into closer correspondence with parametric oscillatory instabilities in optomechanical systems~\cite{braginsky2001parametric,kippenberg2005analysis}, where parametric gain must overcome dissipation for the instability to occur. More broadly, independently controllable signed-frequency pseudomodes could enlarge the class of bath correlation functions and spectral densities that can be represented efficiently in trapped-ion simulations of open quantum systems. This possibility is motivated by pseudomode and thermofield constructions, in which continuum or finite-temperature environments are represented by auxiliary discrete bosonic modes, together with experimental demonstrations of programmable structured reservoirs in trapped-ion systems~\cite{devega2015thermofield,tamascelli2018nonmarkovian,lambert2019unphysical,pleasance2020pseudomodes,wang2024simulating,sun2025quantum,so2025multimode}.

\vspace{3 em}

\section*{Acknowledgments}

We thank Nina P. Inman for helpful discussions, and Baptiste Debecker, John Martin, and François Damanet for the idea of using collective quadratures, for carefully reading the manuscript, and for providing valuable insights. G.P. acknowledges support from the Welch Foundation Award (grant no. C-2154), the NSF CAREER Award (grant no. PHY-2144910), the NSF NRT Award (grant no. 2346014), and the Office of Naval Research (grant no. N00014-23-1-2665). We acknowledge that this material is based on work supported by the U.S. Department of Energy, Office of Science, Office of Nuclear Physics under the Early Career Award (grant no. DE-SC0023806). H.P. acknowledges support from the NSF (grant no. PHY-2513089) and the Welch Foundation (grant no. C-1669).

\bibliography{ref}

\appendix

\section{Negative-frequency quantum Rabi model}
\label{app:neg_rabi}

The negative-frequency quantum Rabi model satisfies
\begin{equation}
H_{\rm QRM}(\omega_0,-\nu,g)=-\sigma_y H_{\rm QRM}(\omega_0,\nu,g)\sigma_y.
\label{eq:negative_positive_qrm_relation}
\end{equation}
Thus, given
\begin{equation}
H_{\rm QRM}(\omega_0,\nu,g)\ket{\psi_E^{(\nu)}}=E\ket{\psi_E^{(\nu)}},
\end{equation}
where $E$ and $\ket{\psi_E^{(\nu)}}$ can be solved analytically \cite{xie2017rabi}, the corresponding eigenvalue equation for the negative-frequency eigenstate is
\begin{equation}
H_{\rm QRM}(\omega_0,-\nu,g)\sigma_y\ket{\psi_E^{(\nu)}}=-E\sigma_y\ket{\psi_E^{(\nu)}}.
\label{eq:specinv}
\end{equation}
The negative-frequency quantum Rabi model therefore has the inverted spectrum of the positive-frequency model, with corresponding eigenstates related by a spin rotation. Equivalently, its closed-system evolution is related to the time-reversed evolution of the positive-frequency counterpart by the same fixed spin transformation,
\begin{equation}
e^{-iH_{\rm QRM}(\omega_0,-\nu,g)t}=\sigma_y e^{+iH_{\rm QRM}(\omega_0,\nu,g)t}\sigma_y.
\end{equation}

Moreover, the $\mathbb{Z}_2$ parity symmetry is preserved for either sign of the effective frequency,
\begin{equation}
[H_{\rm QRM}(\omega_0,\pm\nu,g),\mathcal P]=0,
\end{equation}
with $\mathcal P=\sigma_z e^{i\pi a^\dagger a}$. However, since $\sigma_y$ anti-commutes with $\mathcal P$, the spectral-inversion mapping also reverses the parity eigenvalue. Thus, a parity-$p$ eigenstate of the positive-frequency model is mapped by $\sigma_y$ to a parity-$(-p)$ eigenstate of the negative-frequency model, such that $(E,p)\rightarrow(-E,-p)$.

\section{Diagnostic tools for spin-boson simulators}
\label{sec:diagnostic}

Let us consider a multimode, multispin Hamiltonian of the form
\begin{eqnarray}
H(\{\nu_k\})=H_{\rm s}&+&\sum_k \nu_k a_k^\dagger a_k \\
&+&\sum_j\sum_k g_{jk}\sigma^{j}_{\phi_{jk}}\left(a_k e^{-i\psi_{jk}}+ a_k^\dagger e^{i\psi_{jk}}\right), \nonumber
\label{eq:diagnostic_general_H}
\end{eqnarray}
which extends Eq.~\eqref{eq:multimode} to multiple spins~\cite{so2026thesis}. Here, $H_{\rm s}$ is the spin Hamiltonian generated by combinations of native trapped-ion spin operations, while $j$ and $k$ label the ion and motional mode addressed by the spin-dependent forces, respectively. The sign of each effective frequency is determined by its corresponding sideband detuning through $\nu_k=-\delta_k$. Reversing the signs of all of these detunings implements $\nu_k\rightarrow-\nu_k$ without changing the trapping conditions or any other parameters of the spin-dependent drives.

Suppose that there exists a spin rotation $R_\pi$ satisfying
\begin{equation}
R_\pi H_{\rm s}R_\pi^\dagger = -H_{\rm s}, \quad R_\pi \sigma^j_{\phi_{jk}}R_\pi^\dagger = -\sigma^j_{\phi_{jk}} \quad \text{for all }j,k.
\label{eq:spin_reversal_condition}
\end{equation}
Generally, the $\pi$ rotation is chosen to reverse spin components perpendicular to its rotation axis but leaves the parallel component unchanged. Exact Hamiltonian reversal therefore requires all relevant spin terms to lie in a plane perpendicular to a common pulse axis, and components along the axis must be removed, phase shifted separately, or treated as controlled imperfections. For the quantum Rabi model in Eq.~\eqref{eq:qrm}, one may choose the spin rotation $R_\pi=\sigma_y$. Since $R_\pi$ commutes with the bosonic operators, simultaneous reversal of the mode frequencies then gives
\begin{equation}
H(\{-\nu_k\}) = -R_\pi H(\{\nu_k\})R_\pi^\dagger.
\label{eq:diagnostic_reversal_condition}
\end{equation}

In a trapped-ion implementation, the required sign changes need not always be produced by explicit spin rotations. For interactions generated by phase-controlled laser fields, shifting an appropriate optical phase by $\pi$ can directly reverse the sign of the corresponding term~\cite{lee2005phasecontrol,haljan2005sdf,monroe2021programmable,pagano2025varenna}. For example, a $\pi$ shift of the spin phase of a spin-dependent force changes $\sigma^j_{\phi_{jk}}\rightarrow -\sigma^j_{\phi_{jk}}$, while a $\pi$ shift of its motional phase changes $a_k e^{-i\psi_{jk}}+a_k^\dagger e^{i\psi_{jk}}\rightarrow -\left(a_k e^{-i\psi_{jk}}+a_k^\dagger e^{i\psi_{jk}}\right)$. Either operation reverses the sign of the associated spin-boson interaction. Similarly, the signs of transverse spin terms generated by carrier drives can also be reversed by shifting their optical phases by $\pi$~\cite{monroe2021programmable,pagano2025varenna}. Thus, when all terms in $H_{\rm s}$ and in the spin-boson coupling are independently phase controllable, one may implement the required sign reversal entirely through optical-phase updates, without inserting a physical spin pulse. This phase-based implementation is, however, control dependent. Terms whose signs are not set by an optical phase, such as static longitudinal fields or detuning-generated $\sigma_z^j$ terms, must instead be reversed by changing the corresponding detuning, by applying an appropriate spin rotation, or by combining both operations.

\begin{figure}
    \centering
    \includegraphics[width=1\linewidth]{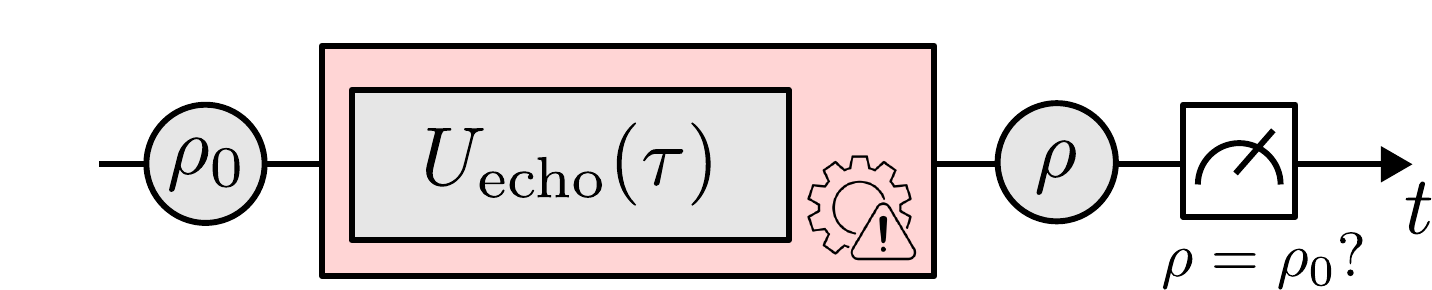}
    \caption{Loschmidt-echo protocol. An input state, typically a pure state $\rho_0=\ket{\psi_0}\bra{\psi_0}$, is subjected to the forward- and reverse-evolution operations that constitute $U_{\rm echo}(\tau)$. The final state $\rho$ is then measured to assess recovery of the initial state. In the absence of drive defects and irreversible processes, the echo is perfect, yielding $\rho=\rho_0$.}
    \label{fig:echo}
\end{figure}

The forward evolution can be undone using the Loschmidt-echo sequence,
\begin{equation}
U_{\rm echo}(\tau)=R_\pi^\dagger e^{-iH(\{-\nu_k\})\tau}R_\pi e^{-iH(\{\nu_k\})\tau}=\identity.
\label{eq:diagnostic_echo}
\end{equation}
When the required spin conjugation can be implemented through the phases of the driven spin operators, the explicit rotations $R_\pi$ and $R_\pi^\dagger$ may instead be absorbed into appropriate optical-phase shifts~\cite{lee2005phasecontrol,haljan2005sdf,monroe2021programmable,pagano2025varenna}. Operationally, the system evolves forward for a duration $\tau$, a spin $\pi$ pulse is applied, all spin-dependent-force detunings are reversed, the system evolves for the same duration, and the inverse spin pulse is applied~\cite{garttner2017otoc,lewisswan2019fotoc}. This sequence, which is summarized in Fig.~\ref{fig:echo}, provides an internal consistency check of the analog simulator. For an ideal closed Hamiltonian satisfying Eq.~\eqref{eq:diagnostic_reversal_condition}, the echo returns the system to its initial state. A residual echo defect reveals Hamiltonian terms or dynamical processes that are not reversed by the protocol, including imperfect detuning or phase control, residual carrier terms, spectator-mode couplings, spin dephasing, motional heating, dissipation, and spin-independent motional nonlinearities~\cite{garttner2017otoc,sanchez2009loschmidt}. Higher-order corrections generated by the spin-dependent force~\cite{pagano2025varenna,so2026thesis} may also not reverse exactly under the echo sequence and can therefore be characterized through their contributions to the echo defect.

Overall, the protocol separates the intended reversible spin-boson dynamics from calibration errors and irreversible channels. In this sense, we remark that the protocol should be regarded as a self-consistency and reversibility benchmark rather than a complete certification of the target Hamiltonian. Nevertheless, it remains useful when the intended dynamics appear to relax or dephase within the accessible observation window, while coherent oscillations or revivals emerge only at much longer times. A successful echo then indicates that the apparent relaxation is predominantly reversible and consistent with unitary dephasing within the spin-boson system, whereas an echo defect indicates control errors or irreversible processes.

\begin{figure}
    \centering
    \includegraphics[width=1\linewidth]{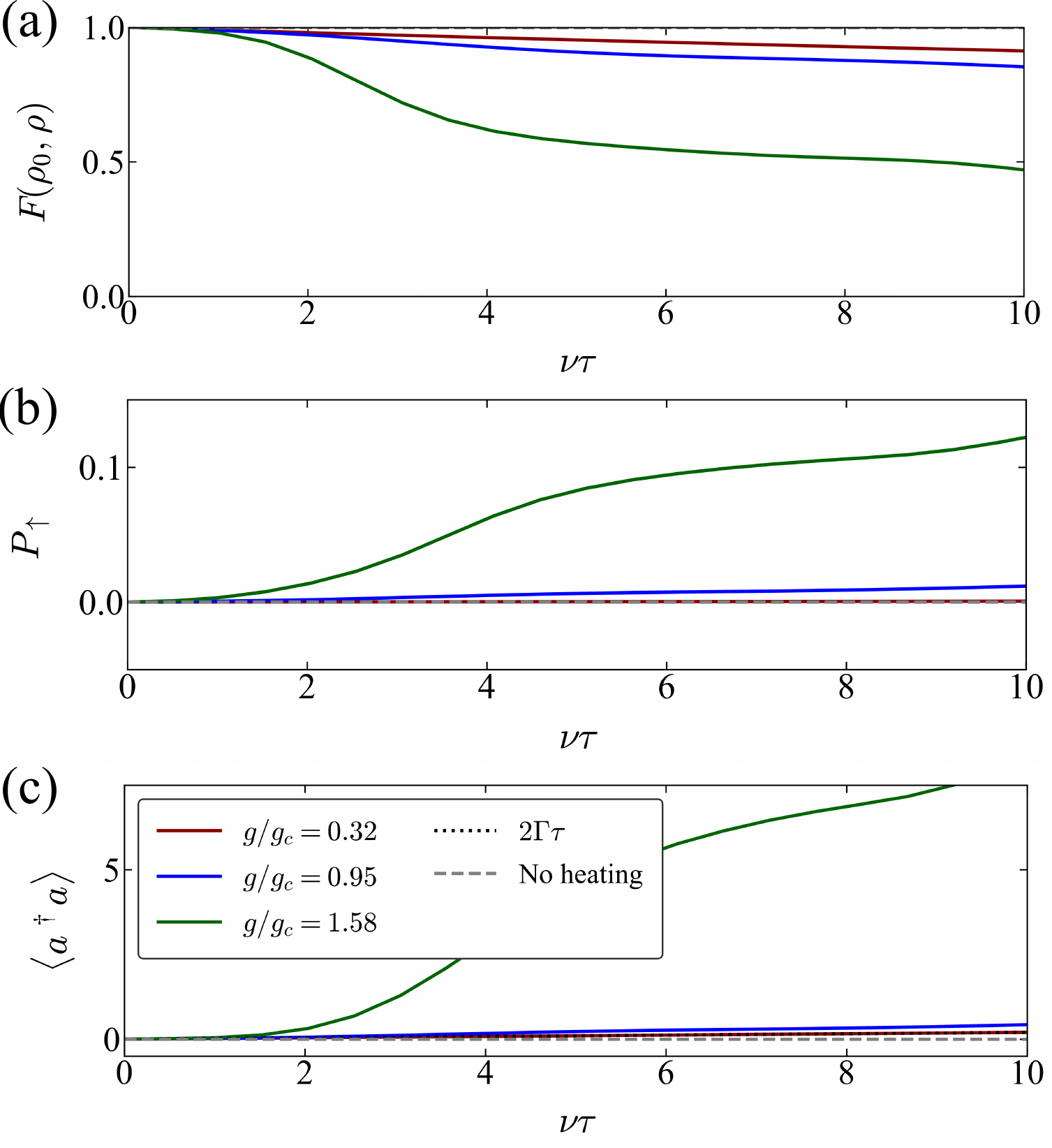}
    \caption{Loschmidt-echo application to the quantum Rabi model subject to motional heating, a common noise process in trapped-ion platforms~\cite{wineland1997issue,myatt2000decoherence}. (a)--(c) Root Uhlmann fidelity, excited-spin population, and mean phonon number as functions of the echo duration $\tau$ in the normal-like ($g=0.32g_c$), near-critical ($g=0.95g_c$), and superradiant-like ($g=1.58g_c$) regimes. Results are shown for the finite frequency ratio $\omega_0/\nu=40$ and motional-heating rate $\Gamma=0.01\nu$, illustrating how heating drives the final state away from the initial state and increasingly suppresses echo recovery near and above the critical coupling $g_c$. Black solid lines indicate perfect recovery of the initial state, $\rho_0=\ket{\downarrow}\bra{\downarrow}\otimes\ket{0}\bra{0}$. In panel (c), the dashed black line shows the bare phonon-number increase due to heating, $2\Gamma\tau$. In the normal-like regime, the post-echo mean phonon number closely follows this baseline and is accompanied by the slight increase in spin excitation shown in panel (b). The bosonic Hilbert space in the numerical simulations is truncated to $N_{\rm cut}=200$ Fock states.}
    \label{fig:1mrabi}
\end{figure}

To illustrate how Loschmidt echoes can diagnose imperfections in spin-boson simulations, we consider a quantum Rabi model subject to motional heating at a rate $\Gamma$. Its open-system dynamics ($\Gamma\neq 0$) are governed by the Lindblad master equation
\begin{align}
\frac{\partial \rho}{\partial t}&=-i[H_{\rm QRM},\rho]+\Gamma\left(\mathcal{D}_{a}[\rho]+\mathcal{D}_{a^\dagger}[\rho]\right).
\label{eq:masterQRM}
\end{align}
In the absence of coherent dynamics ($[H_{\rm QRM},\rho]=0$), the equal-rate dissipators $\mathcal{D}_{a}$ and $\mathcal{D}_{a^\dagger}$ lead to $d\braket{ a^\dagger a}/dt=\Gamma$, so that over the total echo duration $2\tau$, the bare mean phonon number increases by $2\Gamma\tau$.

In the limit $\omega_0/\nu\rightarrow\infty$, the closed quantum Rabi model exhibits a superradiant quantum phase transition at the critical spin-boson coupling $g_c=\frac{1}{2}\sqrt{\omega_0\nu}$. At finite $\omega_0/\nu$, this transition is replaced by a crossover between normal-like and superradiant-like regimes. Fig.~\ref{fig:1mrabi} shows the recovery of the initial state
\begin{equation}
\rho_0=\ket{\downarrow,0}\bra{\downarrow,0}=\ket{\downarrow}\bra{\downarrow}\otimes\ket{0}\bra{0},
\end{equation}
after a variable-duration Loschmidt-echo sequence $U_{\rm echo}(\tau)$, in which the forward and reverse evolution intervals each have duration $\tau$. We take $\Gamma=0.01\nu$ and $\omega_0=40\nu$, and consider several values of $g$ spanning the normal-like, near-critical, and superradiant-like regimes. We quantify the recovery using the root Uhlmann fidelity
\begin{equation}
F(\rho_0,\rho)=\operatorname{Tr}\sqrt{\sqrt{\rho_0}\rho\sqrt{\rho_0}}=\sqrt{\bra{\downarrow,0}\rho\ket{\downarrow,0}},
\end{equation}
the excited-spin population $P_\uparrow=\operatorname{Tr}\left[\left(\ket{\uparrow}\bra{\uparrow}\otimes\identity_{\rm b}\right)\rho\right]$, and the mean phonon number $\braket{a^\dagger a}=\operatorname{Tr}\left[\left(\identity_{\rm s}\otimes a^\dagger a\right)\rho\right]$ with $\rho$ denoting the state at the end of the complete echo sequence. 

In the absence of heating ($\Gamma=0$), ideal Hamiltonian reversal gives $F(\rho_0,\rho)=1$, $P_\uparrow=0$, and $\braket{a^\dagger a}=0$ for all $g/g_c$ (indicated by black solid lines). We find that bosonic heating increasingly degrades the echo near and above the critical coupling, $g\gtrsim g_c$, where enhanced bosonic fluctuations and stronger spin-boson correlations amplify the effect of irreversible motional diffusion. By contrast, in the normal-like regime, the initial state remains close to the weakly dressed spin-boson vacuum and undergoes comparatively little coherent excitation. The resulting small phonon occupation and weak spin-boson correlations make the echo less sensitive to motional heating. This observation also provides insight into the robustness of the motional dynamics in the opposite-sign two-mode Rabi model discussed in the main text, whose stable, critical, and dynamically unstable regimes are all described relative to the vacuum-centered normal configuration rather than a stabilized superradiant phase.

In addition to benchmarking the performance of the analog spin-boson simulators, combining signed-frequency reversal with local spin rotations or motional displacements enables measurements of out-of-time-order correlators (OTOCs). For unitary operators $V$ and $W$, the conventional OTOC is
\begin{equation}
F(t)=\braket{W^\dagger(t)V^\dagger W(t)V},\quad W(t)=e^{iHt}We^{-iHt},
\label{eq:diagnostic_otoc}
\end{equation}
where the expectation value is taken with respect to the chosen initial state~\cite{garttner2017otoc,swingle2016scrambling,garttner2018entanglement}. The probe operators $V$ and $W$ may be chosen as spin rotations, motional displacements, phonon-parity operations, or joint spin-phonon drives. By implementing the required backward evolution, the signed-frequency echo extends time-reversal-based OTOC protocols developed for spin-only systems~\cite{garttner2017otoc,swingle2016scrambling,garttner2018entanglement} to spin-boson simulators with nontrivial bosonic phase-space dynamics.

A fidelity OTOC is obtained by replacing one of the operators with the initial-state projector $V=\rho_0=\ket{\psi_0}\bra{\psi_0}$ and choosing $W=e^{i\delta\phi G}$, where $G$ is a bosonic quadrature. This quantity has been studied in the quantum Rabi and Dicke models as probes of scrambling, spin-phonon entanglement, thermalization, and quantum chaos~\cite{kirkova2022otoc,lewisswan2019fotoc}. Related quadrature-based commutator OTOCs have also been investigated in the Dicke model and its variants~\cite{tiwari2023dicke}.

\section{Position-dependent spin rotation}
\label{app:position_rotation}

By adapting the position-dependent spin rotation of Ref.~\cite{zhu2026dissipative}, we provide an alternative derivation of the dispersive interaction. We again write the two-mode quantum Rabi Hamiltonian as $H_{\rm 2QRM}=H_0+V$ with $V=\sigma_x\widetilde{X}=\sigma_x\left[g_a(a+a^\dagger)+g_b(b+b^\dagger)\right]$. When the energy scale of the spin is much larger than the bosonic frequency ($\omega_0\gg|\nu_k|$), we treat the position operator in $V$ as a quasi-static coordinate such that the spin-dependent part of $H_{\rm 2QRM}$ becomes a two-level Hamiltonian with
longitudinal splitting $\omega_0$ and transverse field
$2\widetilde{X}$. We introduce the position-dependent rotation
\begin{equation}
U(\widetilde{X})=\exp\left[-\frac{i}{2}\sigma_y\theta(\widetilde{X})\right],
\quad \theta(\widetilde{X})=\arctan\left(\frac{2\widetilde{X}}{\omega_0}\right).
\label{eq:position_rotation}
\end{equation}
Using $U^\dagger\sigma_zU=\sigma_z\cos\theta-\sigma_x\sin\theta$, $U^\dagger\sigma_xU=\sigma_x\cos\theta+\sigma_z\sin\theta$, and $\tan\theta=2\widetilde{X}/\omega_0$ gives
\begin{equation}
U^\dagger\left(\frac{\omega_0}{2}\sigma_z+\sigma_x\widetilde{X}\right)U=\frac{\sigma_z}{2}\sqrt{\omega_0^2+4\widetilde{X}^2}.
\label{eq:exact_spin_diagonalization}
\end{equation}

To evaluate the transformation of the pure bosonic terms $H_B=\nu_a a^\dagger a + \nu_b b^\dagger b$, we introduce the quadratures
\begin{equation}
x_k=\frac{k+k^\dagger}{\sqrt{2}},\quad
p_k=\frac{k-k^\dagger}{i\sqrt{2}},
\label{eq:app_quadratures}
\end{equation}
such that
\begin{equation}
\widetilde{X}=\sqrt{2}\sum_{k=a,b}g_kx_k,\quad
H_B=\sum_{k=a,b}\frac{\nu_k}{2}\left(x_k^2+p_k^2-1\right).
\label{eq:app_HB_quadratures}
\end{equation}
Since $U$ depends only on the position quadratures, we obtain
\begin{equation}
U^\dagger x_kU=x_k,\quad
U^\dagger p_kU=p_k-\mathcal A_k(\widetilde{X})\sigma_y,
\label{eq:transformed_quadratures}
\end{equation}
where we have defined the spin-dependent gauge potential
\begin{equation}
\mathcal A_k(\widetilde{X})\equiv\frac{1}{2}\frac{\partial\theta}{\partial x_k}=\frac{\sqrt{2}\,g_k\omega_0}{\omega_0^2+4\widetilde{X}^2}.
\label{eq:gauge_potential}
\end{equation}
Equivalently, the ladder operators transform as
\begin{equation}
U^\dagger kU=k-i\epsilon_k(\widetilde{X})\sigma_y,\quad
\epsilon_k(\widetilde{X})=\frac{\mathcal A_k(\widetilde{X})}{\sqrt{2}}.
\label{eq:transformed_ladder}
\end{equation}

Therefore, the fully transformed two-mode quantum Rabi Hamiltonian is
\begin{eqnarray}
\widetilde{H}
&=& U^\dagger H_{\rm 2QRM}U\nonumber\\
&=&\frac{\sigma_z}{2}\sqrt{\omega_0^2+4\widetilde{X}^2}\\
&\quad&\,+\sum_{k=a,b}\frac{\nu_k}{2}\left[x_k^2+\left(p_k-\mathcal{A}_k\sigma_y\right)^2-1\right]\nonumber
\label{eq:exact_rotated_H}
\end{eqnarray}
Expanding the kinetic contribution gives
\begin{eqnarray}
\widetilde{H}&=&\frac{\sigma_z}{2}
\sqrt{\omega_0^2+4\widetilde{X}^2}+\nu_a a^\dagger a + \nu_b b^\dagger b \label{eq:rotated_H_expanded}\\
&\quad&-\frac{\sigma_y}{2}\sum_{k=a,b}\nu_k\left\{p_k,\mathcal A_k\right\}+\frac{1}{2}\sum_{k=a,b}\nu_k\mathcal A_k^2.\nonumber
\end{eqnarray}
The fourth term in Eq.~\eqref{eq:rotated_H_expanded} is
off-diagonal in the dressed-spin basis, whereas the fifth
term is spin-independent. For $\|\widetilde{X}\|/\omega_0\ll1$, the effective gauge potential has the expansion
\begin{equation}
\mathcal A_k=\frac{\sqrt{2}\,g_k}{\omega_0} + \mathcal{O}\left(\frac{g_k\|\widetilde{X}\|^2}{\omega_0^3}\right).
\label{eq:gauge_expansion}
\end{equation}
Consequently, the operator factors entering the off-diagonal
and spin-independent contributions above become
\begin{eqnarray}
\sum_{k=a,b}\nu_k\left\{p_k,\mathcal A_k\right\}
&=&\frac{2\sqrt{2}}{\omega_0}\sum_{k=a,b}\nu_kg_kp_k\label{eq:leading_offdiagonal_term}\\
&\quad&\,+\,\mathcal{O}\left(\sum_{k=a,b}\frac{|\nu_kg_k|}{\omega_0^3}\left\|\left\{p_k,\widetilde{X}^2\right\}\right\|\right),
\nonumber
\end{eqnarray}
\begin{equation}
    \sum_{k=a,b}\nu_k\mathcal A_k^2=\sum_{k=a,b}\frac{2\nu_kg_k^2}{\omega_0^2}+\mathcal{O}\left(\sum_{k=a,b}\frac{|\nu_k|g_k^2\|\widetilde{X}\|^2}{\omega_0^4}\right).
\label{eq:leading_geometric_correction}
\end{equation}
In the dispersive regime and with $|\nu_k|\ll\omega_0$, we can omit these correction terms, resulting in the adiabatic Hamiltonian
\begin{equation}
    H_{\rm ad}= \frac{\sigma_z}{2}\sqrt{\omega_0^2+4\widetilde{X}^2}+\nu_a a^\dagger a + \nu_b b^\dagger b.
\label{eq:nonlinear_adiabatic_H}
\end{equation}
Expanding the dressed-spin term gives
\begin{equation}
    \frac{\sigma_z}{2}\sqrt{\omega_0^2+4\widetilde{X}^2}=\frac{\omega_0}{2}\sigma_z + \frac{\sigma_z}{\omega_0}\widetilde{X}^2 + \mathcal{O}\left(\frac{\|\widetilde{X}\|^4}{\omega_0^3}\right).
\end{equation}
Retaining the quadratic contribution reproduces the effective
Hamiltonian in Eq.~\eqref{eq:2qrmSW}. Although the position-dependent rotation treatment is nonperturbative in $\widetilde{X}/\omega_0$ for the spin-dependent part and identifies the nonlinear dressed-spin potential, neglecting the residual terms generated by $U^\dagger (\nu_a a^\dagger a + \nu_b b^\dagger b) U$ remains an adiabatic approximation. Within the common regime of validity, this approach is therefore complementary to, rather than more rigorous than, the Schrieffer-Wolff expansion used in the main text.

\section{Instability threshold}
\label{app:pair_amplification}

In the following, we determine the instability threshold of the dynamics generated by the effective Hamiltonian in Eq.~\eqref{eq:detuned_pair_hamiltonian}. The corresponding Heisenberg equations are
\begin{align}
\frac{d a}{dt}&=i\left[H_{\rm RWA}^{(s)},a\right]=-i\widetilde{\nu}_a a-is\lambda b^\dagger,\\
\frac{d b^\dagger}{dt}&=i\left[H_{\rm RWA}^{(s)},b^\dagger\right]=i\widetilde{\nu}_b b^\dagger+is\lambda a.
\end{align}
These equations can be written in matrix form as
\begin{equation}
\frac{d}{dt}
\begin{pmatrix}
a\\
b^\dagger
\end{pmatrix}
=\left[-\frac{i}{2}\left(\widetilde{\nu}_a-\widetilde{\nu}_b\right)\identity+\mathcal{M}\right]
\begin{pmatrix}
a\\
b^\dagger
\end{pmatrix},
\label{eq:app_pair_matrix_equation}
\end{equation}
where
\begin{equation}
\mathcal{M}=\begin{pmatrix}
-\dfrac{i(\widetilde{\nu}_a+\widetilde{\nu}_b)}{2}&-is\lambda\\
is\lambda&\dfrac{i(\widetilde{\nu}_a+\widetilde{\nu}_b)}{2}
\end{pmatrix},
\end{equation}
whose eigenvalues are
\begin{equation}
    m_{\pm}=\pm\sqrt{\lambda^2-\frac{\Delta\widetilde{\nu}^2}{4}}.
\end{equation}
The first term in Eq.~\eqref{eq:app_pair_matrix_equation} produces only a common phase rotation and therefore does not affect the bosonic occupations. The nontrivial dynamics is determined by
\begin{equation}
\mathcal{M}^2=\left(\lambda^2-\frac{\Delta\widetilde{\nu}^2}{4}\right)\identity.
\label{eq:app_matrix_square}
\end{equation}
First, consider the condition $|\lambda|>|\Delta\widetilde{\nu}|/2$ in which the eigenvalues $m_{\pm}$ are real. We define the positive instability growth rate as
\begin{equation}
\Gamma_{\rm inst}=\sqrt{\lambda^2-\frac{\Delta\widetilde{\nu}^2}{4}}.
\label{eq:app_instability_rate}
\end{equation}
The two dynamical solutions consequently contain factors $e^{\pm\Gamma_{\rm inst}t}$, corresponding to exponentially growing and decaying collective modes.

The formal solution to Eq.~\eqref{eq:app_pair_matrix_equation} can be factorized as
\begin{equation}
\begin{pmatrix}
a(t)\\
b^\dagger(t)
\end{pmatrix}
=e^{-i(\widetilde{\nu}_a-\widetilde{\nu}_b)t/2}e^{\mathcal{M}t}
\begin{pmatrix}
a(0)\\
b^\dagger(0)
\end{pmatrix}.
\label{eq:app_formal_matrix_solution}
\end{equation}
By using Eq.~\eqref{eq:app_matrix_square} and separating the matrix exponential into even and odd powers of $\mathcal{M}$, we obtain
\begin{align}
e^{\mathcal{M}t}&=\sum_{n=0}^{\infty}\frac{(\mathcal{M}t)^n}{n!}
\nonumber\\
&=\cosh\left(\Gamma_{\rm inst}t\right)\identity
+\frac{\sinh\left(\Gamma_{\rm inst}t\right)}
{\Gamma_{\rm inst}}\mathcal{M}.
\label{eq:app_matrix_exponential}
\end{align}
The corresponding Bogoliubov transformation is~\cite{caves1985twophoton1,schumaker1985twophoton2}
\begin{align}
a(t)=&\bigg[\left(\cosh\left(\Gamma_{\rm inst}t\right)-i\frac{\Delta\widetilde{\nu}}{2\Gamma_{\rm inst}}\sinh\left(\Gamma_{\rm inst}t\right)\right)a(0)
\nonumber\\
&\quad-is\frac{\lambda}{\Gamma_{\rm inst}}\sinh\left(\Gamma_{\rm inst}t\right) b^\dagger(0)\bigg]e^{-i(\widetilde{\nu}_a-\widetilde{\nu}_b)t/2},
\label{eq:app_detuned_bogoliubov_a}
\end{align}
\begin{align}
b(t)=&\bigg[\left(\cosh\left(\Gamma_{\rm inst}t\right)-i\frac{\Delta\widetilde{\nu}}{2\Gamma_{\rm inst}}\sinh\left(\Gamma_{\rm inst}t\right)\right)b(0)
\nonumber\\
&\quad-is\frac{\lambda}{\Gamma_{\rm inst}}\sinh\left(\Gamma_{\rm inst}t\right)
a^\dagger(0)\bigg]e^{+i(\widetilde{\nu}_a-\widetilde{\nu}_b)t/2}.
\label{eq:app_detuned_bogoliubov_b}
\end{align}
These transformations preserve the bosonic commutation relations~\cite{caves1985twophoton1,schumaker1985twophoton2}.

\begin{figure}
    \centering
    \includegraphics[width=1\linewidth]{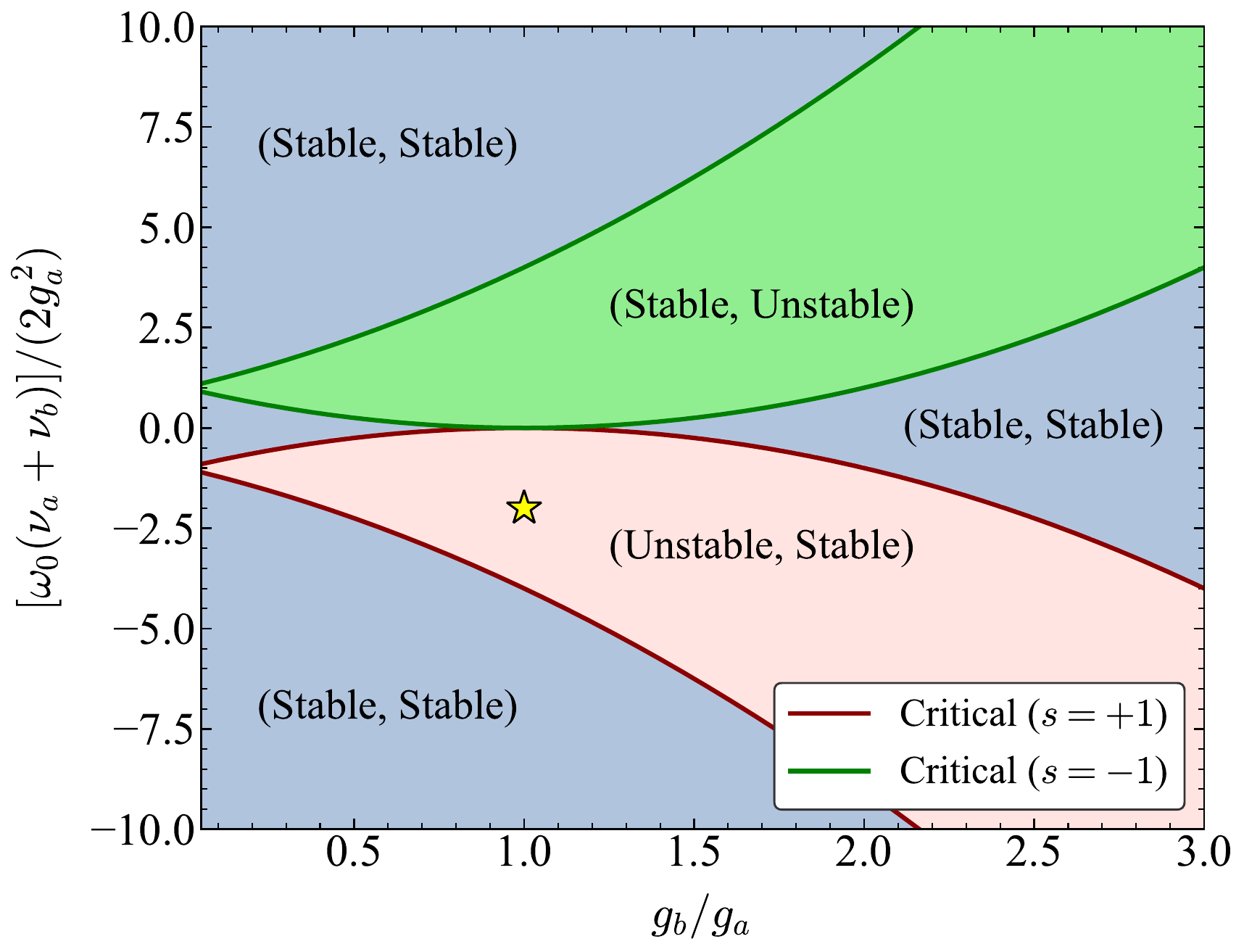}
    \caption{Dynamical stability diagram predicted by the effective Hamiltonian as a function of the normalized bare frequency mismatch and coupling ratio for both spin branches, relevant to the discussion in Sec.~\ref{sec:spincontrol}. In each region, the stability classifications of the $s=\pm1$ branches are listed in the order $(\text{regime for }s=+1,\text{ regime for }s=-1)$. The colored solid curves mark the critical conditions for the corresponding spin branches. The yellow star highlights the parameter point $(1,-2)$ used in Figs.~\ref{fig:spincontrol} and \ref{fig:spincontrol_logneg} of the main text, where the $s=+1$ branch is dynamically unstable while the $s=-1$ branch remains stable. Although not all blue regions are individually labeled, they correspond to regimes in which both spin branches are dynamically stable.}
    \label{fig:stabilitydiagram}
\end{figure}

For an initial two-mode vacuum, $\ket{\psi(0)}=\ket{0_a,0_b}$, the two mode occupations are therefore
\begin{equation}
\braket{n_a(t)}=\braket{n_b(t)}=\frac{\lambda^2}{\Gamma_{\rm inst}^2}\sinh^2\left(\Gamma_{\rm inst}t\right).
\label{eq:app_detuned_pair_growth}
\end{equation}
Thus, they grow exponentially at long times,
\begin{equation}
\braket{n_k(t)}\sim\frac{\lambda^2}{4\Gamma_{\rm inst}^2}e^{2\Gamma_{\rm inst}t}, \quad k=a,b,
\end{equation}
demonstrating the dynamical instability of the detuned pair Hamiltonian~\cite{jain2007quantum,kohler2018negative}. On exact dressed resonance, $\Delta\widetilde{\nu} = 0$, $\Gamma_{\rm inst}=|\lambda|$, and Eq.~\eqref{eq:app_detuned_pair_growth} reduces to Eq.~\eqref{eq:resonance_phonon_growth}. Thus, the resonant pair interaction also produces exponential amplification at long times, $\braket{n_k(t)}\sim e^{2|\lambda| t}/4$. On the other hand, below the instability threshold, $|\lambda|<|\Delta\widetilde{\nu}|/2$, we define
\begin{equation}
\Omega_{\rm st}=\sqrt{\frac{\Delta\widetilde{\nu}^2}{4}-\lambda^2}.
\label{eq:app_pair_oscillation_frequency}
\end{equation}
In this case, the eigenvalues of $\mathcal{M}$ are purely imaginary, and the vacuum occupations undergo bounded oscillations following
\begin{equation}
\braket{n_k(t)}=\frac{\lambda^2}{\Omega_{\rm st}^2}\sin^2\left(\Omega_{\rm st}t\right).
\label{eq:app_stable_pair_dynamics}
\end{equation}
At the critical boundary, $|\lambda|=|\Delta\widetilde{\nu}|/2$, we have $\mathcal{M}^2=0$, so that $e^{\mathcal{M}t}=\identity+\mathcal{M}t$.
Equivalently, taking the limit
$\Gamma_{\rm inst}\rightarrow0$ in
Eq.~\eqref{eq:app_detuned_pair_growth} gives
\begin{equation}
\braket{n_k(t)}=\lambda^2t^2.
\label{eq:app_critical_pair_growth}
\end{equation}
The critical dynamics is therefore neither exponentially unstable nor bounded and oscillatory; instead, the occupations grow algebraically. Fig.~\ref{fig:stabilitydiagram} shows the dynamical stability diagram determined by the analytical stability thresholds of the effective Hamiltonian for both spin branches, as a function of the normalized bare frequency mismatch and the spin-boson coupling ratio.

Finally, both the free and pair-interaction terms conserve the bosonic number difference $\left[H_{\rm RWA}^{(s)},n_a-n_b\right]=0$. As a consequence, evolutions from the two-mode vacuum support only correlated Fock states of the form $\ket{n_a=n,n_b=n}$. Pair amplification can therefore increase the total bosonic occupation while preserving $n_a-n_b=0$ for every populated Fock-state component, as expected for a two-mode squeezed vacuum~\cite{caves1985twophoton1,schumaker1985twophoton2}.

\section{Schrieffer-Wolff frame in the numerical calculations}
\label{app:sw_frame_numerics}

To compare the full two-mode quantum Rabi model directly with the analytical predictions of the effective quadratic model, we express the full dynamics in the Schrieffer-Wolff frame. By defining $U_{\rm SW}=e^S$, the dressed full Hamiltonian is
\begin{equation}
H_{\rm 2QRM}^{(\rm d)}=U_{\rm SW}H_{\rm 2QRM}U_{\rm SW}^\dagger.
\end{equation}
We initialize the dressed system in the state $\rho_0$, corresponding to the chosen spin state and the two-mode vacuum, and evaluate its dynamics using the same canonical effective-frame observables $O$ as in the quadratic model.

Numerically, we implement this procedure by evolving the bare
Hamiltonian from the initial state
\begin{equation}
\rho_{\rm bare}(0)=U_{\rm SW}^\dagger\rho_0U_{\rm SW}.
\label{eq:bare_initial_state}
\end{equation}
Under the full two-mode quantum Rabi Hamiltonian, this state evolves as
\begin{eqnarray}
\rho_{\rm bare}(t)&=&e^{-iH_{\rm 2QRM}t}\rho_{\rm bare}(0)e^{iH_{\rm 2QRM}t}
\nonumber\\
&=&e^{-iH_{\rm 2QRM}t}U_{\rm SW}^\dagger\rho_0U_{\rm SW}e^{iH_{\rm 2QRM}t}.
\label{eq:bare_evolved_state}
\end{eqnarray}
Transforming the evolved state back to the Schrieffer-Wolff frame then gives
\begin{eqnarray}
\rho_{\rm d}(t)&=&U_{\rm SW}\rho_{\rm bare}(t)U_{\rm SW}^\dagger\nonumber\\
&=&U_{\rm SW}e^{-iH_{\rm 2QRM}t}U_{\rm SW}^\dagger\rho_0 U_{\rm SW}e^{iH_{\rm 2QRM}t}U_{\rm SW}^\dagger\nonumber\\
&=&e^{-iH_{\rm 2QRM}^{(\rm d)}t}\rho_0e^{iH_{\rm 2QRM}^{(\rm d)}t}.
\label{eq:dressed_evolved_state}
\end{eqnarray}
The reported dressed-frame expectation values are thus
\begin{equation}
\langle O(t)\rangle=\operatorname{Tr}\!\left[\rho_{\rm d}(t)O\right].
\end{equation}

For the parameters considered throughout this work, the corrections associated with the Schrieffer-Wolff dressing are small within the bosonic occupation sector satisfying the dispersive condition. Consequently, neglecting the dressing of the initial state and observables and directly comparing the bare-frame dynamics of the full two-mode quantum Rabi model with the leading-order quadratic predictions yields nearly identical results and does not affect our conclusions. This observation is particularly relevant to Sec.~\ref{sec:spincontrol}, where the spin is initialized in a coherent superposition. All numerical simulations in this work were performed using QuTiP~\cite{johansson2013qutip}. Unless otherwise specified, each bosonic mode is truncated to $N_{\rm cut}=400$ Fock states.

\section{Two-mode squeezing}
\label{app:squeeze}

Here, we show explicitly that, within a fixed spin sector, the effective pair dynamics of the opposite-sign two-mode quantum Rabi model for an initial two-mode vacuum generate a two-mode-squeezed state. By using the Bogoliubov transformations for $a(t)$ and $b(t)$ in Eqs.~\eqref{eq:app_detuned_bogoliubov_a} and \eqref{eq:app_detuned_bogoliubov_b}, the anomalous pair correlation for an initial two-mode vacuum is
\begin{eqnarray}
\braket{a(t)b(t)}&=&\left[\cosh\left(\Gamma_{\rm inst}t\right)-i\frac{\Delta\widetilde{\nu}}{2\Gamma_{\rm inst}}\sinh\left(\Gamma_{\rm inst}t\right)\right]\nonumber\\
&\quad&\times\left[-is\frac{\lambda}{\Gamma_{\rm inst}}\sinh\left(\Gamma_{\rm inst}t\right)\right].
\label{eq:app_squeezing_pair_correlation}
\end{eqnarray}
Let us define the mean occupation of either mode as $N(t) \equiv \braket{n_k(t)} = (\lambda^2/\Gamma_{\rm inst}^2)\sinh^2\left(\Gamma_{\rm inst}t\right)$. The magnitude of the pair correlation is therefore
\begin{equation}
\left|\braket{a(t)b(t)}\right|=\sqrt{N(t)\left[N(t)+1\right]},
\label{eq:app_squeezing_pair_correlation_magnitude}
\end{equation}
which saturates the Cauchy-Schwarz bound given by
\begin{equation}
\left|\braket{a(t)b(t)}\right|^2\leq\braket{a(t)a^\dagger(t)}\braket{b^\dagger(t)b(t)}=N(t)\left[N(t)+1\right].
\end{equation}
Thus, the anomalous pair coherence is maximal for the given mode occupation. As shown in Appendix~\ref{app:pair_amplification}, evolution from the two-mode vacuum remains in the correlated-pair subspace spanned by $\ket{n,n}$.

To show that these correlations correspond to two-mode squeezing, we introduce the phase-dependent quadratures of mode $k$
\begin{equation}
    X_k(\phi)=\frac{k e^{-i\phi}+k^\dagger e^{i\phi}}{\sqrt{2}}, \quad P_k(\phi)=X_k\left(\phi+\frac{\pi}{2}\right),
\label{eq:app_squeezing_quadrature}
\end{equation}
for which the vacuum variance is $\bra{0_a,0_b}\Delta X_k^2(\phi)\ket{0_a,0_b}$ $=\bra{0_a,0_b}\Delta P_k^2(\phi)\ket{0_a,0_b}=1/2$. Since the phase of the pair correlation determines which collective quadratures are squeezed, we define it as
\begin{equation}
\varphi(t)=\arg\!\left[\braket{a(t)b(t)}\right],
\end{equation}
where $\braket{a(t)b(t)}=\left|\braket{a(t)b(t)}\right|e^{i\varphi(t)}$. For an arbitrary reference phase $\phi$, we introduce the commuting collective quadratures
\begin{eqnarray}
    X&=&\frac{1}{\sqrt{2}}\left[X_a(\phi)+X_b(\varphi+\pi-\phi)\right],\\
    \label{eq:app_squeezing_x}
    P&=&\frac{1}{\sqrt{2}}\left[X_a\left(\phi+\frac{\pi}{2}\right)-X_b\left(\varphi+\pi-\phi+\frac{\pi}{2}\right)\right],\qquad
    \label{eq:app_squeezing_p}
\end{eqnarray}
such that $[X,P]=0$. Their variances are
\begin{equation}
    \braket{\Delta X^2}=\braket{\Delta P^2}=\frac{1}{2}\left[2N(t)+1-2\sqrt{N(t)[N(t)+1]}\right].
\end{equation}
It is useful to express the variances through an effective squeezing parameter, $r(t)\equiv{\rm arsinh}{\sqrt{N(t)}}$, yielding
\begin{equation}
\braket{\Delta X^2}=\braket{\Delta P^2}=\frac{1}{2}e^{-2r(t)},
\label{eq:app_squeezing_variance}
\end{equation}
where we have used $\cosh x-\sinh x=e^{-x}$. The corresponding orthogonal collective quadratures are then amplified, $\braket{\Delta X_{\perp}^2}=\braket{\Delta P_{\perp}^2}=\frac{1}{2}e^{+2r(t)}$.
Thus, although neither mode is individually quadrature squeezed, the joint quadratures are squeezed below their two-mode-vacuum variance. In fact, the reduced state of each mode is thermal, with a phase-independent quadrature variance
\begin{equation}
    \braket{\Delta X_a^2(\phi)}=\braket{\Delta X_b^2(\phi)}=N(t)+\frac{1}{2}=
    \frac{1}{2}\cosh\left[2r(t)\right].
\label{eq:app_squeezing_local_variance}
\end{equation}
The squeezing therefore resides entirely in correlations between the two modes.

On exact dressed pair resonance, $\Delta\widetilde{\nu}=0$, the effective squeezing parameter simplifies to $r(t)=|\lambda|t$, and thus the evolved bosonic state can be written explicitly as
\begin{equation}
    \ket{\psi(t)}=\frac{1}{\cosh\left(|\lambda|t\right)}\sum_{n=0}^{\infty}\left[-is\,\operatorname{sgn}(\lambda)\tanh\left(|\lambda|t\right)\right]^n\ket{n,n}.
\label{eq:app_squeezing_tmsv}
\end{equation}
This is a two-mode-squeezed vacuum whose squeezing phase depends on the spin sector $s$ and on the sign of $\lambda$. The phase determines which collective quadratures are squeezed but does not change the squeezing magnitude or bosonic populations.

Away from exact resonance, the squeezing behavior follows directly from the
occupation dynamics. In the unstable regime,
\begin{equation}
r(t)=\operatorname{arsinh}\left[\frac{|\lambda|}{\Gamma_{\rm inst}}\sinh\left(\Gamma_{\rm inst}t\right)\right].
\label{eq:app_squeezing_unstable_r}
\end{equation}
At long times, it grows asymptotically as
\begin{equation}
r(t)=\Gamma_{\rm inst}t+\ln\left(\frac{|\lambda|}{\Gamma_{\rm inst}}\right)+\mathcal{O}\left(e^{-2\Gamma_{\rm inst}t}\right),
\label{eq:app_squeezing_unstable_asymptotic}
\end{equation}
so that the variance decreases exponentially as $e^{-2\Gamma_{\rm inst}t}$. At the critical boundary, $|\lambda|=|\Delta\widetilde{\nu}|/2$, one has
\begin{equation}
r(t)=\operatorname{arsinh}\left(|\lambda|t\right).
\label{eq:app_squeezing_critical_r}
\end{equation}
The pair population grows quadratically, whereas the squeezing parameter grows logarithmically at long times. Lastly, in the stable regime, the effective squeezing parameter is
\begin{equation}
r(t)=\operatorname{arsinh}\left[\frac{|\lambda|}{\Omega_{\rm st}}\left|\sin\left(\Omega_{\rm st}t\right)\right|\right].
\label{eq:app_squeezing_stable_r}
\end{equation}
Therefore, the system still produces transient two-mode squeezing even below the instability threshold, but the squeezing periodically returns to zero rather than increasing without bound.

\section{Higher-order dispersive corrections}
\label{app:higher_order_SW}

The effective Hamiltonian in Eq.~\eqref{eq:2qrmSW} is obtained by retaining the leading nontrivial terms of the Schrieffer-Wolff expansion in the dispersive regime. The subsequent rotating-wave approximation then isolates the resonant pair-creation dynamics discussed in the main text. To separate the effects of the rotating-wave approximation from those associated with neglected higher-order dispersive corrections, we additionally compare the full two-mode Rabi dynamics directly with the dynamics generated by Eq.~\eqref{eq:2qrmSW}, without applying the rotating-wave approximation.

\begin{figure}
    \centering
    \includegraphics[width=1\linewidth]{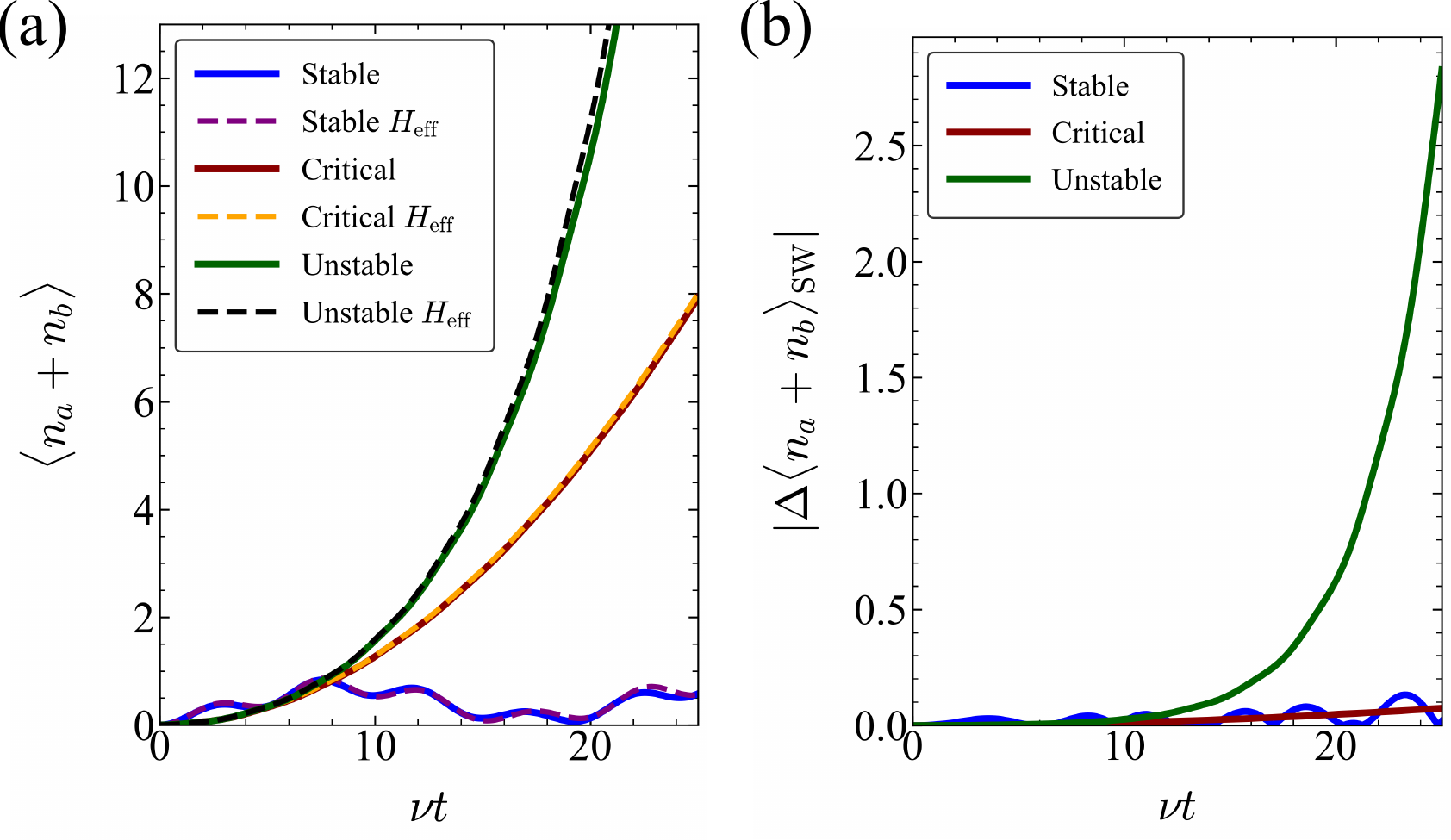}
    \caption{Effects of higher-order dispersive corrections. (a) Total bosonic population, $\braket{n_a+n_b}$, in the stable, critical, and unstable regimes, using the same parameters as in Fig.~\ref{fig:paircreation}(b). Solid curves show the full two-mode quantum Rabi dynamics transformed to the Schrieffer-Wolff frame, while dashed curves show the dynamics generated by the effective Hamiltonian in Eq.~\eqref{eq:2qrmSW} before applying the rotating-wave approximation. (b) Corresponding absolute deviation in the total bosonic population between the two dynamics.}
    \label{fig:app_SW_comparison}
\end{figure}

\begin{figure*}
    \centering
    \includegraphics[width=1\linewidth]{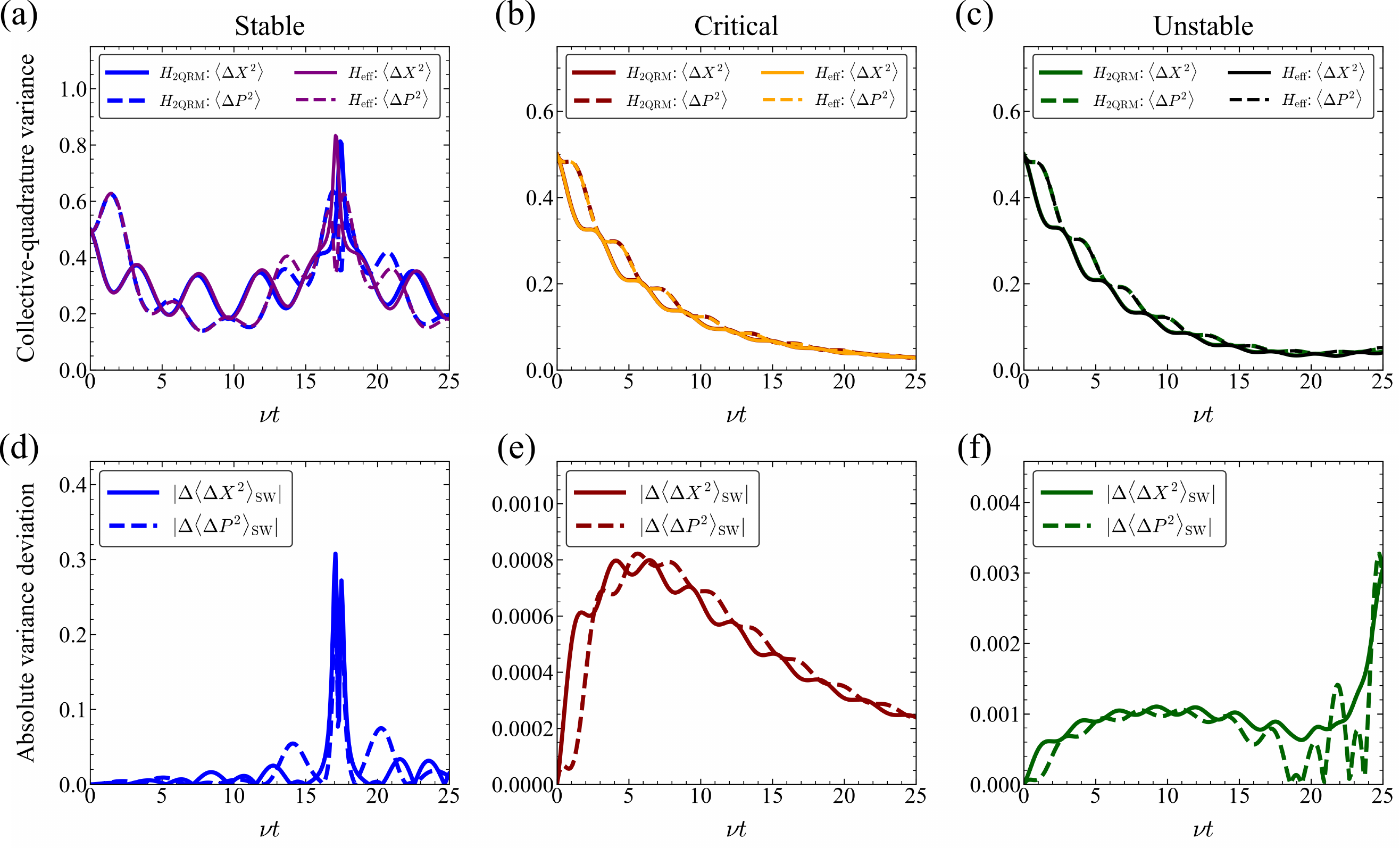}
    \caption{Higher-order dispersive corrections to collective-quadrature squeezing. (a--c) Collective-quadrature variances in the stable, critical, and dynamically unstable regimes, respectively, using the same parameters as in Fig.~\ref{fig:squeeze}. Solid and dashed curves denote $\langle\Delta X^2\rangle$ and $\langle\Delta P^2\rangle$, respectively. Colored curves show the full two-mode Rabi dynamics transformed to the Schrieffer-Wolff frame, while the corresponding comparison curves show the dynamics generated by the effective Hamiltonian in Eq.~\eqref{eq:2qrmSW} before applying the rotating-wave approximation. (d--f) Corresponding absolute deviations in $\langle\Delta X^2\rangle$ and $\langle\Delta P^2\rangle$ between the two descriptions.}
    \label{fig:app_SW_squeezing}
\end{figure*}

As shown in Fig.~\ref{fig:app_SW_comparison}, this comparison is performed for the same stable, critical, and dynamically unstable parameter sets considered in Fig.~\ref{fig:paircreation}(b). The dynamics generated by Eq.~\eqref{eq:2qrmSW} closely follow those of the full two-mode Rabi model over the simulated time window, particularly in the stable and critical regimes. The remaining deviations are small, indicating that corrections beyond the retained Schrieffer-Wolff order are weak for the parameters considered. In particular, these deviations are substantially smaller than those between the quadratic Schrieffer-Wolff dynamics and the pair-creation rotating-wave approximation shown in Fig.~\ref{fig:paircreation}(b), indicating that the dominant quantitative discrepancy in the effective description arises from the rotating-wave approximation rather than from higher-order dispersive corrections. At longer times in the unstable regime, however, the rapidly growing bosonic occupation gradually weakens the dispersive condition, leading to the increasing deviation visible in Fig.~\ref{fig:app_SW_comparison}(b), although the deviation remains small over the simulated time window.

For completeness, we also perform the same comparison for the collective-quadrature variances underlying the squeezing shown in Fig.~\ref{fig:squeeze}. As shown in Fig.~\ref{fig:app_SW_squeezing}, the full two-mode Rabi dynamics and the dynamics generated by Eq.~\eqref{eq:2qrmSW} without the rotating-wave approximation are still in close agreement. Thus, the same conclusion that higher-order dispersive corrections remain small over the timescales considered applies.

\section{Logarithmic negativity in a bipartite spin-boson system}
\label{sec:lognegder}

In this section, we derive Eq.~\eqref{eq:spin_boson_logneg_reduced} from Eq.~\eqref{eq:spin_boson_logneg}, using the fact that the global spin-boson state remains pure under unitary evolution. Since the spin subsystem is two-dimensional, the Schmidt decomposition of $\ket{\Psi(t)}$ contains at most two nonzero terms~\cite{ekert1995entangled} and is written as
\begin{equation}
    \ket{\Psi(t)}=\sqrt{c_+(t)}\ket{+}_{\rm s}\ket{+}_{\rm b}+\sqrt{c_-(t)}\ket{-}_{\rm s}\ket{-}_{\rm b},
\end{equation}
where $c_\pm(t)\geq 0$, $c_+(t)+c_-(t)=1$, and the spin and bosonic
Schmidt states form orthonormal sets. The corresponding density matrix is
\begin{align}
    \rho(t)&=c_+(t)\ket{+}_{\rm s}\ket{+}_{\rm b}\bra{+}_{\rm s}\bra{+}_{\rm b}\nonumber\\
    &\quad\,+\,c_-(t)\ket{-}_{\rm s}\ket{-}_{\rm b}\bra{-}_{\rm s}\bra{-}_{\rm b}\nonumber\\
    &\quad\,+\,\sqrt{c_+(t)c_-(t)}\ket{+}_{\rm s}\ket{+}_{\rm b}\bra{-}_{\rm s}\bra{-}_{\rm b}\nonumber\\
    &\quad\,+\,\sqrt{c_+(t)c_-(t)}\ket{-}_{\rm s}\ket{-}_{\rm b}\bra{+}_{\rm s}\bra{+}_{\rm b}.
\end{align}
Taking the partial transpose with respect to the spin subsystem gives
\begin{align}
    \rho^{T_{\rm s}}(t)&=c_+(t)\ket{+}_{\rm s}\ket{+}_{\rm b}\bra{+}_{\rm s}\bra{+}_{\rm b}\nonumber\\
    &\quad\,+\,c_-(t)\ket{-}_{\rm s}\ket{-}_{\rm b}\bra{-}_{\rm s}\bra{-}_{\rm b}\nonumber\\
    &\quad\,+\,\sqrt{c_+(t)c_-(t)}\ket{-}_{\rm s}\ket{+}_{\rm b}\bra{+}_{\rm s}\bra{-}_{\rm b}\nonumber\\
    &\quad\,+\,\sqrt{c_+(t)c_-(t)}\ket{+}_{\rm s}\ket{-}_{\rm b}\bra{-}_{\rm s}\bra{+}_{\rm b}.
\end{align}
Therefore,
\begin{align}
    \left\|\rho^{T_{\rm s}}(t)\right\|_1&=c_+(t)+c_-(t)+2\sqrt{c_+(t)c_-(t)}\nonumber\\
    &=1+2\sqrt{c_+(t)c_-(t)}.
    \label{eq:partialtranspose}
\end{align}
Since $c_\pm(t)$ are the eigenvalues of the reduced-spin density matrix given by Eq.~\eqref{eq:reduced_spin_density}, we have
\begin{align}
c_+(t)c_-(t)&=\det\rho_{\rm s}(t)\nonumber\\
&=P_\uparrow(t)P_\downarrow(t)-\left|\rho_{\uparrow\downarrow}(t)\right|^2.
\end{align}
By substituting this result into Eq.~\eqref{eq:partialtranspose} and then into Eq.~\eqref{eq:spin_boson_logneg}, we obtain Eq.~\eqref{eq:spin_boson_logneg_reduced}.

\section{Analytical logarithmic negativities in the quadratic model}
\label{app:analytic_logneg}

Analytical expressions for the bipartite logarithmic negativities discussed in Sec.~\ref{sec:spincontrol} can be obtained within the dispersive rotating-wave approximation. For the initial state in Eq.~\eqref{eq:coherent_initial}, the effective Hamiltonian in Eq.~\eqref{eq:hamiltonian_effective_spincontrol} gives the entangled state in Eq.~\eqref{eq:conditional_state}. As discussed in Appendix~\ref{app:pair_amplification}, $H_{\rm RWA}^{(s)}$ conserves $n_a - n_b$. Therefore, evolution from the two-mode vacuum remains in the subspace spanned by $\ket{n,n}$, and the number-difference term in Eq.~\eqref{eq:detuned_pair_hamiltonian} does not affect the conditional bosonic states.

For a fixed spin sector $s=\pm 1$, the Bogoliubov transformation derived in Eqs.~\eqref{eq:app_detuned_bogoliubov_a} and \eqref{eq:app_detuned_bogoliubov_b} can be written as
\begin{equation}
a(t)=\left[u_s(t)a(0)+v_s(t)b^\dagger(0)\right]e^{-i(\tilde{\nu}_a-\tilde{\nu}_b)t/2},
\label{eq:app_bog_a}
\end{equation}
where, in the dynamically unstable regime,
\begin{align}
u_s(t)&=
\cosh\!\left[\Gamma_{\rm inst}(s)t\right]-i\frac{\Delta\tilde{\nu}(s)}
{2\Gamma_{\rm inst}(s)}\sinh\!\left[\Gamma_{\rm inst}(s)t\right],
\label{eq:app_us}\\
v_s(t)&=-is\frac{\lambda}{\Gamma_{\rm inst}(s)}\sinh\!\left[\Gamma_{\rm inst}(s)t\right],
\label{eq:app_vs}
\end{align}
with $\Gamma_{\rm inst}(s)=\sqrt{\lambda^2-\Delta\tilde{\nu}^2(s)/4}$. The coefficients satisfy $|u_s(t)|^2-|v_s(t)|^2=1$. In the stable regime,
$|\Delta\tilde{\nu}(s)|>2|\lambda|$, the corresponding
expressions for $u_s(t)$ and $v_s(t)$ are 
\begin{align}
u_s(t)&=\cos\!\left[\Omega_{\rm st}(s)t\right]-i\frac{\Delta\tilde{\nu}(s)}{2\Omega_{\rm st}(s)}\sin\!\left[\Omega_{\rm st}(s)t\right],
\label{eq:app_us_stable}\\
v_s(t)&=-is\frac{\lambda}{\Omega_{\rm st}(s)}\sin\!\left[\Omega_{\rm st}(s)t\right],
\label{eq:app_vs_stable}
\end{align}
with $\Omega_{\rm st}(s)=\sqrt{\Delta\tilde{\nu}^2(s)/4-\lambda^2}$. The critical solution follows from the corresponding
$\Gamma_{\rm inst}(s)\rightarrow0$ limit. 

By expanding the conditional state as $\ket{\phi_s(t)}=\sum_{n=0}^{\infty}c_n^{(s)}(t)\ket{n,n}$, we obtain the recurrence $c_{n+1}^{(s)}(t)=[v_s(t)/u_s^*(t)]c_n^{(s)}(t)$ from the Bogoliubov relation. This fixes the relative Fock-state amplitudes but leaves the overall phase of the conditional state undetermined. After normalization, each conditional state is a two-mode-squeezed vacuum of the form
\begin{equation}
\ket{\phi_s(t)} = e^{i\vartheta_s(t)}\sqrt{1-|\zeta_s(t)|^2}\sum_{n=0}^{\infty}\zeta_s^n(t)\ket{n,n},
\label{eq:app_conditional_tmsv}
\end{equation}
where $\zeta_s(t)\equiv v_s(t)/u_s^*(t)$. The vacuum component of the Schr\"odinger equation, $i\dot{c}_0^{(s)}(t)=s\lambda c_1^{(s)}(t)$, with $c_0^{(s)}(0)=1$, gives
\begin{equation}
c_0^{(s)}(t)=\frac{e^{i\Delta\widetilde{\nu}(s)t/2}}{u_s^*(t)}.
\end{equation}
Consequently, the phase in the conditional state satisfies
\begin{equation}
e^{i\vartheta_s(t)}=e^{i\Delta\widetilde{\nu}(s)t/2}\frac{|u_s(t)|}{u_s^*(t)}.
\end{equation}
Although this overall phase does not affect observables within an individual spin branch, the relative phase between the two branches must be retained in their overlap and in spin postselection. The overlap between the two spin-conditioned bosonic states is
\begin{align}
\Phi(t)&= e^{i[\vartheta_{+1}(t)-\vartheta_{-1}(t)]} \nonumber\\
&\quad\times\frac{\sqrt{(1-|\zeta_{-1}(t)|^2)(1-|\zeta_{+1}(t)|^2)}}{1-\zeta_{-1}^*(t)\zeta_{+1}(t)},
\label{eq:app_phi}
\end{align}
where we have used $\sum_{n=0}^\infty x^n=1/(1-x)$ for $|x|< 1$. Since $1-|\zeta_s|^2=1/|u_s|^2$, we get
\begin{equation}
|\Phi(t)|=\frac{1}{\left|u_{-1}(t)u_{+1}^*(t)-v_{-1}^*(t)v_{+1}(t)\right|}.
\label{eq:app_phi_uv}
\end{equation}
Since $|\Phi(t)|=\mathcal{C}(t)$ for nonzero initial spin amplitudes, we can substitute Eq.~\eqref{eq:app_phi_uv} into Eq.~\eqref{eq:exp_logneg_sab} of the main text to obtain an analytic expression for the logarithmic negativity across the $s|(a,b)$ bipartition within the dispersive rotating-wave approximation.

The remaining two bipartitions can also be treated analytically within the quadratic theory. Since the global state in Eq.~\eqref{eq:conditional_state} remains pure, the logarithmic negativity across the bipartition $a|(s,b)$ is determined entirely by the eigenvalues of the reduced state of mode $a$. By using Eq.~\eqref{eq:app_conditional_tmsv}, tracing over the spin and mode $b$ leads to then
\begin{equation}
\rho_a(t)=\sum_{n=0}^{\infty}p_n(t)|n\rangle\langle n|,
\label{eq:app_rho_a}
\end{equation}
where
\begin{align}
p_n(t)={}&|\alpha|^2\left[1-|\zeta_{+1}(t)|^2\right]|\zeta_{+1}(t)|^{2n}\nonumber\\
&+|\beta|^2\left[1-|\zeta_{-1}(t)|^2\right]|\zeta_{-1}(t)|^{2n}.
\label{eq:app_pn}
\end{align}
The absence of interference terms follows from the orthogonality of the two spin states when the spin is traced out. In terms of the conditional mean occupations $N_s(t)=|v_s(t)|^2$ and by using $|\zeta_s(t)|^2=N_s(t)/[N_s(t)+1]$, Eq.~\eqref{eq:app_pn} becomes
\begin{align}
p_n(t)={}&|\alpha|^2\frac{N_{+1}^n(t)}{\left[N_{+1}(t)+1\right]^{n+1}}+
|\beta|^2\frac{N_{-1}^n(t)}{\left[N_{-1}(t)+1\right]^{n+1}}.
\label{eq:app_pn_N}
\end{align}

For a globally pure bipartite state with reduced-state eigenvalues $\{p_n\}$, the trace norm of the partial transpose is~\cite{vidal2002computable}
\begin{equation}
\left\|\rho^{T_a}\right\|_1=\left(\sum_{n=0}^{\infty}\sqrt{p_n}\right)^2.
\end{equation}
The logarithmic negativity across $a|(s,b)$ is therefore
\begin{equation}
E_{\mathcal N}^{a|(s,b)}(t)=2\log_2\left[\sum_{n=0}^{\infty}\sqrt{p_n(t)}\right],
\label{eq:app_logneg_a}
\end{equation}
with $p_n(t)$ given by Eq.~\eqref{eq:app_pn_N}. Since the quadratic pair dynamics conserve $n_a-n_b$ and, starting from the two-mode vacuum, populate only states $\ket{n,n}$, we have $\rho_b(t)=\rho_a(t)$, and thus an identical result is obtained for the $b|(s,a)$ bipartition, $E_{\mathcal N}^{b|(s,a)}(t)=E_{\mathcal N}^{a|(s,b)}(t)$.

\section{Entanglement of spin-postselected non-Gaussian bosonic states}
\label{app:postselected_non_gaussian}

We note that the non-Gaussian bosonic states obtained by spin postselection can remain entangled even when their entanglement is not detected by the second-moment Duan-Simon criterion. As discussed in the main text, the bosonic state conditioned on the $\pm$ outcome remains within the $n_a-n_b=0$ subspace and can be written in the normalized form
\begin{equation}
\ket{\psi_{\pm,\theta}(t)}=\sum_{n=0}^{\infty}d_n^{(\pm,\theta)}(t)\ket{n,n},
\quad\sum_{n=0}^{\infty}\left|d_n^{(\pm,\theta)}(t)\right|^2= 1.
\label{eq:app_postselected_schmidt}
\end{equation}
The coefficients $d_n^{(\pm,\theta)}(t)$ are determined by the spin-analysis basis and the two spin-conditioned squeezing parameters through Eqs.~\eqref{eq:postselected_bosonic_state_general} and \eqref{eq:postselected_cn}. After absorbing the phases of $d_n^{(\pm,\theta)}(t)$ into a local number-state basis, Eq.~\eqref{eq:app_postselected_schmidt} is a Schmidt decomposition with Schmidt coefficients $\left|d_n^{(\pm,\theta)}(t)\right|$. The corresponding mean occupation of either bosonic mode is
\begin{equation}
\braket{a^\dagger a}=\braket{b^\dagger b}\equiv N_{\pm,\theta}(t)=\sum_{n=0}^{\infty}n\left|d_n^{(\pm,\theta)}(t)\right|^2,
\label{eq:app_postselected_N}
\end{equation}
and the anomalous pair correlation is
\begin{equation}
M_{\pm,\theta}(t)\equiv\braket{ab}=\sum_{n=1}^{\infty}n\,\left[d_{n-1}^{(\pm,\theta)}(t)\right]^*d_n^{(\pm,\theta)}(t).
\label{eq:app_postselected_M}
\end{equation}
By using the phase-rotated collective quadratures introduced in the main text, we obtain
\begin{align}
\braket{\Delta X^2}=\braket{\Delta P^2}=&N_{\pm,\theta}(t)+\frac{1}{2}\nonumber\\&\quad+{\rm Re}\left[e^{-i(\phi_a+\phi_b)}M_{\pm,\theta}(t)\right].
\label{eq:app_postselected_varX}
\end{align}
The corresponding Duan-Simon quantity is therefore
\begin{equation}
\mathcal{V}_{\rm DS}^{(\pm,\theta)}(t)=2N_{\pm,\theta}(t)+1+2\,{\rm Re}\left[e^{-i(\phi_a+\phi_b)}M_{\pm,\theta}(t)\right].
\label{eq:app_postselected_DS}
\end{equation}
The minimum is attained for quadrature phases satisfying $\phi_a+\phi_b=\arg[M_{\pm,\theta}(t)]+\pi$, yielding
\begin{equation}
\mathcal{V}_{\rm DS,min}^{(\pm,\theta)}(t)=2N_{\pm,\theta}(t)+1-2\left|M_{\pm,\theta}(t)\right|.
\label{eq:app_postselected_DS_min}
\end{equation}
For Gaussian states, entanglement can be fully characterized by second-order moments through appropriate covariance-matrix criteria~\cite{duan2000inseparability,simon2000peres}. In the present quadrature normalization, a Duan-Simon violation occurs when $\left|M_{\pm,\theta}(t)\right|>N_{\pm,\theta}(t)$. For the non-Gaussian postselected states considered here, however, this condition is sufficient but not necessary for entanglement~\cite{hertz2016detection}. This can be seen directly from the Schmidt structure of Eq.~\eqref{eq:app_postselected_schmidt}. For a bipartite pure state, separability is equivalent to having Schmidt rank one~\cite{horodecki2009quantum}. Therefore, the two bosonic modes are entangled whenever more than one Schmidt coefficient $\left|d_n^{(\pm,\theta)}(t)\right|$ is nonzero. For a globally pure bipartite state, the trace norm of the partial transpose is
\begin{equation}
\left\|\rho^{T_a}\right\|_1=\left[\sum_{n=0}^{\infty}\left|d_n^{(\pm,\theta)}(t)\right|\right]^2,
\end{equation}
and hence its logarithmic negativity is
\begin{equation}
E_{\mathcal N}^{a|b,(\pm,\theta)}(t)=2\log_2\left(\sum_{n=0}^{\infty}\left|d_n^{(\pm,\theta)}(t)\right|\right).
\label{eq:app_pure_logneg_general}
\end{equation}
Here, we have used the pure-state relation between the Schmidt coefficients and the trace norm of the partial transpose~\cite{vidal2002computable,plenio2005logarithmic}. Therefore, the logarithmic negativity can remain finite even when $\mathcal{V}_{\rm DS}^{(\pm,\theta)}(t)\geq 1$.

We now derive the results for the illustrative even-pair construction introduced in the main text. We take $\zeta_{+1}=\zeta$, $\zeta_{-1}=-\zeta$, and $|\alpha|=|\beta|=1/\sqrt{2}$, and choose the spin-analysis phase so that the two contributions to the $+$ postselection outcome have equal overall coefficients. The coefficient of $|n,n\rangle$ is then proportional to
\begin{equation}
\zeta^n+(-\zeta)^n=\left[1+(-1)^n\right]\zeta^n,
\end{equation}
resulting in the cancellation of all odd pair-number components. After normalization, the postselected state is given by Eq.~\eqref{eq:main_even_pair_state}. Comparing with the general Schmidt expansion in Eq.~\eqref{eq:app_postselected_schmidt}, the corresponding Schmidt amplitudes are
\begin{equation}
d_{2m}^{(\rm even)}=\sqrt{1-|\zeta|^4}\,\zeta^{2m},\quad d_{2m+1}^{(\rm even)}=0.
\label{eq:app_even_pair_dn}
\end{equation}
For any $0<|\zeta|<1$, multiple Schmidt coefficients are nonzero, and the two bosonic modes are therefore entangled.

At the same time, the anomalous second-order correlation vanishes,
\begin{equation}
M_{\rm even}=\langle ab\rangle=0,
\label{eq:app_even_pair_Mzero}
\end{equation}
because the operator $ab$ connects states $\ket{n,n}$ to states $\ket{n-1,n-1}$, whereas Eq.~\eqref{eq:main_even_pair_state} contains only even values of $n$. The mean occupation of either bosonic mode is
\begin{align}
N_{\rm even} &= (1-|\zeta|^4) \sum_{m=0}^{\infty} 2m\,|\zeta|^{4m}=\frac{2|\zeta|^4}{1-|\zeta|^4}.
\label{eq:app_even_pair_N}
\end{align}
Eq.~\eqref{eq:app_postselected_DS_min} therefore gives
\begin{equation}
\mathcal{V}_{\rm DS,min}^{(\rm even)}=\frac{1+3|\zeta|^4}{1-|\zeta|^4}>1,
\label{eq:app_even_pair_DS}
\end{equation}
for $0<|\zeta|<1$. Thus, this entangled state does not violate the Duan-Simon inequality for any choice of the quadrature phases.

On the other hand, Eq.~\eqref{eq:app_pure_logneg_general} gives
\begin{align}
E_{\mathcal N}^{a|b,(\rm even)}&=
2\log_2\left(\sqrt{1-|\zeta|^4}\sum_{m=0}^{\infty}|\zeta|^{2m}\right)\nonumber\\
&=2\log_2\left(\frac{\sqrt{1-|\zeta|^4}}{1-|\zeta|^2}\right)\nonumber\\
&=\log_2\left(\frac{1+|\zeta|^2}{1-|\zeta|^2}\right).
\label{eq:app_even_pair_logneg}
\end{align}
Thus, the logarithmic negativity is finite for $0<|\zeta|<1$. Consequently, the state is entangled even though it never violates the Duan-Simon inequality. The example demonstrates explicitly how spin postselection can transfer entanglement from Gaussian second-order correlations into higher-order, non-Gaussian correlations that are not captured by the covariance matrix.

\end{document}